\documentclass[showpacs,preprintnumbers,pre,preprint]{revtex4}

\usepackage{graphicx}
\usepackage{amsmath}
\usepackage{amsmath}
\usepackage{amssymb}
\usepackage{psfrag}
\usepackage{color}

\begin{document}

\title{Stability and roughness of tensile cracks in disordered materials}
\author{E. Katzav$^1$ and M. Adda-Bedia$^2$}
\affiliation{
$^1$Department of Mathematics, King's College London, Strand, London WC2R 2LS, United Kingdom\\
$^2$Laboratoire de Physique Statistique, Ecole Normale Sup\'erieure, UPMC Paris 6, Universit\'e Paris Diderot, CNRS, 24 rue Lhomond, 75005 Paris, France}
\email{eytan.katzav@kcl.ac.uk, adda@lps.ens.fr}
\date{\today}

\begin{abstract}
We study the stability and roughness of propagating cracks in heterogeneous brittle two-dimensional elastic materials. We begin by deriving an equation of motion describing the dynamics of such a crack in the framework of Linear Elastic Fracture Mechanics, based on the Griffith criterion and the Principle of Local Symmetry. This result allows us to extend the stability analysis of Cotterell and Rice to disordered materials. In the stable regime we find stochastic crack paths. Using tools of statistical physics we obtain the power spectrum of these paths and their probability distribution function, and conclude they do not exhibit self-affinity. We show that a real-space fractal analysis of these paths can lead to the wrong conclusion that the paths are self-affine. To complete the picture, we unravel the systematic bias in such real-space methods, and thus contribute to the general discussion of reliability of self-affine measurements. 
\end{abstract}


\maketitle

\section{Introduction}

In material science, fractography is concerned with the description of fracture surfaces in solids and is routinely used to determine the cause of failure in engineering structures~\cite{Hull}. Various types of crack growth (fatigue, stress corrosion, cracking...) produce different characteristic features on the surface, which in turn can be used to identify the failure mode and direction. Fractography is one of the most used experimental techniques to recover some aspects of crack dynamics, and is thus a major tool to develop and evaluate theoretical models of crack growth behavior. 

Fractography in two dimensional and three dimensional materials is fundamentally different. A broken surface in three dimensions is the trace of a front line singularity and unless a full dynamical measurement is available, it is not possible to reconstruct its propagation history~\cite{Ravi97,Bonamy12}. An extreme example is the family of systems in which a crack front is confined to move along a weak plane~\cite{Daguier95,Schmittbuhl97,Delaplace99,Maloy06,Chopin} and where the post-mortem surface is simply a flat surface that cannot reveal any information on the actual advance of the crack front. In contrast, a crack path in two dimensions is simply the trace left by the propagation of a crack tip and thus can be thought of as the world-line of a moving singularity. This means that a post-mortem fractographic analysis can fully recover the crack propagation history. This shows that the 2D problem provides a good framework with many advantages to decipher the crack tip dynamics, and is therefore our focus here.

For slowly propagating cracks in homogeneous materials, fracture surfaces are smooth and a fundamental question that arises in that context is the stability of the propagating crack with respect to a prescribed path. A stability analysis of two-dimensional cracks propagating in homogeneous materials based on Linear Elastic Fracture Mechanics (LEFM) was performed by Cotterell and Rice~\cite{Cotterell80}, and yields the famous $T$-criterion. This criterion states that if the quantity called the $T$-stress (see Eq.~(\ref{eq:SIF}) below for a definition) is positive the crack path becomes unstable, whereas if $T \le 0$ the path is stable. Experimentally and theoretically, the instability predicted by the $T$-criterion has been proven to be a necessary condition but not a sufficient one~\cite{Radon77,Yuse93,Marder94,adda95,thermique,Deegan03,Menouillard}.

The stability analysis of Cotterell and Rice being incomplete already in homogeneous media cannot be expected to describe correctly crack propagation in heterogeneous media where the path of the fracture is generally rough. Actually, for such materials, fracture surfaces are claimed to exhibit fractal (or self-affine) properties~\cite{Mandelbrot84,Bouchaud90,Maloy92,Marder97}. The self-affinity of a $d$-dimensional surface is fully characterized by $(d-1)$ exponents~\cite{Schmittbuhl95a,RafiFrac,Oystein}. Since the dynamics of cracks in heterogeneous media is a rich field encompassing a wide range of physical phenomena, it is important to distinguish between three different exponents~: the one describing roughness in the direction perpendicular to the crack propagation~\cite{Ponson06}, the second one describing the roughness in the direction of the propagation (the so called "out-of-the-plane" roughness, which is the subject of this paper) and the third one describing the in-plane roughness of the crack front during its propagation through the material~\cite{Daguier95,Schmittbuhl97,Delaplace99,Maloy06}. In some cases these exponents are related \cite{ineq1,ineq2} but generically they are independent.

There are some experimental measurements of the roughness exponent $\zeta$ of two-dimensional or quasi two-dimensional cracks (in the appendix some methods of measurements of the "out-of-the-plane" roughness exponent are described). For Berea sandstone ($\zeta \simeq 0.8$~\cite{Plouraboue96}); for concrete ($\zeta \simeq 0.75$~\cite{Balankin05}); for paper ($\zeta \simeq 0.6$~\cite{Salminen03}, $\zeta \simeq 0.66$~\cite{Proc06} and $\zeta \simeq 0.7$~\cite{Mallick06}); for wood ($\zeta\simeq 0.68$~\cite{Wood}) and for a system of drinking straws ($\zeta\simeq 0.73$~\cite{Straws}). As can be seen, all the values vary between $0.6$ and $0.8$ and thus suggest non-universal behaviour. In particular note that the measured exponents are all larger than $0.5$, a value that corresponds to the roughness exponent of an uncorrelated random walk~\cite{barabasi95}.

In this work we aim at a thorough study of crack propagation in 2D disordered materials. A natural question is whether the roughness of the broken surface is related to an instability mechanism of the crack tip propagation. In the following we will derive an equation of motion for a crack propagating in disordered medium that would allow us to study both its stability and roughness properties. The paper is organized as follows: we start by recalling the stability analysis \`{a} la Cotterell and Rice. Then we present the formulation of a stochastic model that takes into account the material heterogeneity and uses results regarding kinked cracks, and extend the $T$-criterion to heterogeneous materials. We then specialize to the case $T=0$ and discuss the roughness of the resulting crack surfaces. Thanks to an exact result of the model in that limit~\cite{EPL07}, we are able to obtain analytically the form of the power spectrum of the paths, and offer an alternative interpretation of experimental results. We conclude by discussing the implication of our result on the methodology of self-similarity analysis by suggesting a new measurement bias that has not been considered previously~\cite{Schmittbuhl95a,RafiFrac,Oystein}.

\section{crack paths in homogeneous materials}

The key ingredient that allows a general discussion of cracks in a brittle material is the fact that the static stress field in the vicinity of the crack tip has the following universal expansion
\begin{equation}
\sigma_{ij}(r,\theta)=\sum_{\ell=1,2}\frac{K_\ell}{\sqrt{2\pi r}}\,\Sigma^{(\ell)}_{ij}(\theta)+T\delta_{ix}\delta_{jx}+O\left(\sqrt{r}\right)\;,
\label{eq:SIF}
\end{equation}
where $(r,\theta)$ are polar coordinates with $r=0$ located at the crack tip, and $\Sigma^{(\ell)}_{ij} (\theta)$ are known functions describing the angular variations of the stress field components~\cite{Broberg,Leblond03}. In this expansion, $K_\ell$ ($\ell=1,2,3$) and $T$ are the Stress Intensity Factors (SIFs) and the nonsingular $T$-stress respectively. This singular behaviour of the stress field justifies the expectation that the crack-tip dynamics could be formulated in terms of the SIFs and the $T$-stress alone.

In a 2D material, well-established criteria for quasi-static crack propagation are the Griffith energy criterion~\cite{Broberg} and the Principle of Local Symmetry (PLS) \cite{Gol,adda99}. This is expressed by the following equations of motion
\begin{eqnarray}
{\cal G} &\equiv & \frac{1}{2\mu}K^2_1 = \Gamma \qquad \qquad \qquad \qquad [{\rm Griffith}] \;, \label{eq:grif}\\
K_2 &=& 0 \qquad \qquad \qquad \qquad \qquad \qquad [{\rm PLS}] \;, 
\label{eq:pls}
\end{eqnarray}
where $\mu$ is the Lam\'e shear coefficient and $\Gamma$ is the fracture energy. Eq.~(\ref{eq:grif}) states that in order to induce crack propagation, the energy release rate ${\cal G}$ must be large enough to create new crack surfaces. Eq.~(\ref{eq:pls}) imposes the symmetry of the stress field in the vicinity of the crack tip, such that it is locally under a pure opening mode. Therefore, the crack path is mainly selected by the PLS while Eq.~(\ref{eq:grif}) controls the intensity of the loading necessary to allow propagation. Other propagation criteria have been proposed in the literature~\cite{Mroz10,Sethna93,Afek05}, notably the maximum energy release rate criterion~\cite{Erdogan63} which states that the crack extends in the direction that maximizes the rate of energy release. However, the PLS has been shown to be the only self-consistent one \cite{Leblond92,adda99,Hakim05,Hakim09}.

A linear stability analysis of quasi-static two-dimensional crack propagation in homogeneous materials based on these concepts has been performed by Cotterell and Rice \cite{Cotterell80}, which gave rise to the $T$-criterion. This criterion states that for $T >0$, a tensile crack propagation becomes unstable with respect to small perturbations around the straight path. Otherwise the straight crack propagation is stable. Experimentally, the $T$-criterion is known to hold, at least for cases with $T>0$, that is when the prediction is that cracks become unstable~\cite{Radon77}. However, even when $T \le 0$ the crack path can become unstable in some situations (for example, the thermal crack problem in which the crack path exhibits an oscillatory instability~\cite{Yuse93,adda95,thermique}).

In addition to the $T$-criterion, Cotterell and Rice predicted that for a semi-infinite straight crack experiencing a sudden local shear perturbation the subsequent crack path $h(x)$ scales as $\sqrt{x}$ in the stable regime, while $h(x) \sim e^x$ in the unstable case. However, the result $h(x) \sim \sqrt{x}$ is only marginally stable (i.e. $h'(x) \sim 1/\sqrt{x} \rightarrow 0$ for large $x$'s)~\cite{RafiFrac}, thus reflecting a limited aspect of stability. This situation calls for a revision of the stability properties of slow cracks, especially in heterogeneous materials.

\section{A Crack tip equation of motion in disordered materials}

\begin{figure}[ht]
\includegraphics[width=8cm]{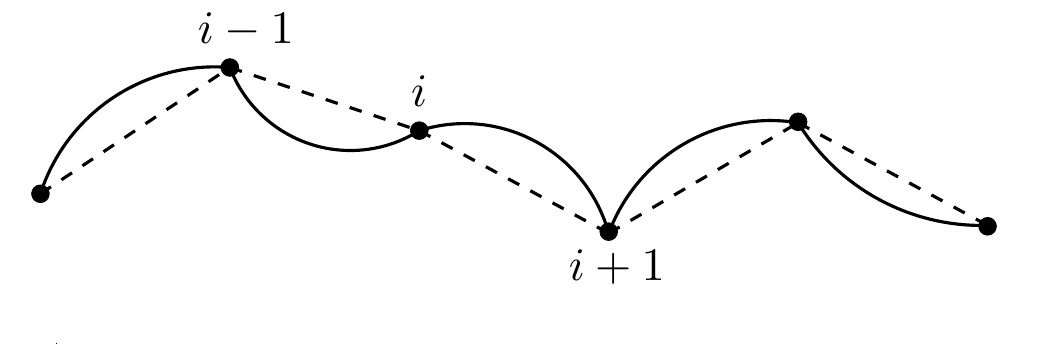}
\caption{Schematic representation a crack path with kinked curved extensions due to encounters with heterogeneities. the index $i$ corresponds to location $x_i$ of the $i^{th}$ heterogeneity.}
\label{fig:curved}
\end{figure}

Based on these observations we propose an equation describing the propagation of a crack in a disordered medium that allows to predict its path and study its stability. Our model is based on a description where all the relevant information is encoded in the SIF's $K_{l}$ and in the $T$-stress. The crack propagation criteria used are the Griffith energy balance (\ref{eq:grif}) and the Principle of Local Symmetry (\ref{eq:pls}). The physical picture of a propagating crack in a disordered material in the current formulation is summarized in Fig.~\ref {fig:curved}. It assumes that the crack tip propagates smoothly until it encounters a heterogeneity that changes locally the fracture energy and induces a local shear perturbation. As a result, the crack forms a kink at a prescribed angle depending on the local perturbation induced by the heterogeneity. In order to calculate this angle it is necessary first to introduce some results regarding kinked cracks.

Consider an elastic body containing a straight crack with a kinked curved extension of length $s$ and kink angle $\theta$. Using standard assumptions related on the scaling of the stress field in the vicinity of the crack tip it is shown that the shape of the local crack extension should be given by~\cite{Leblond89}
\begin{equation}
\label{eq:shape} Y=bX+aX^{3/2}+O(X^2)\;,
\end{equation}
where $b=\tan\theta$ is the slope of the kink and the curvature parameter $a$ quantifies the curved extension of the kink. Moreover, it is shown that the static SIFs at the crack tip after kinking $K^{\prime}_\ell(s)$ $(\ell=1,2)$ are related to the SIFs before kinking $K_\ell$ and to the $T$-stress via~\cite{Leblond89}
\begin{equation}
K^{\prime}_\ell(s) = \sum_{m=1,2}F_{\ell m}(\theta)K_m +
\sum_{m=1,2}\left[G_m(\theta)T\delta_{\ell m}+aH_{\ell
m}(\theta)K_m\right]\sqrt{s}+O(s)\; , \label{eq:exp}
\end{equation}
where $F_{\ell m}$, $G_\ell$ and $H_{\ell m}$ are universal functions in the sense that they do not depend on the geometry of the body nor on the applied loading. They depend only on the kink angle $\theta$ and  were computed in~\cite{Leblond92}. Note that Eq.~(\ref{eq:exp}) shows that in general $K'_2 \neq 0$ unless a special symmetry sets it to zero. Therefore, applying the Principle of Local Symmetry means that the expansion~(\ref{eq:exp}) should vanish order by order in $s$. In the presence of a small shear loading ($\left|K_2\right|\ll K_1$) the extension of the initial straight crack must therefore satisfy
\begin{eqnarray}
b &\simeq& \theta \simeq -2\frac{K_2}{K_1}\;, \label{eq:kink0} \\
a &\simeq& \frac{8}{3}\sqrt{\frac{2}{\pi}} \theta \frac{T}{K_1} \;,
\label{eq:kink1}
\end{eqnarray}
where the expansions of the functions $F_{\ell m}$, $G_{\ell}$ and $H_{\ell m}$ for small angles have been used \cite{Leblond92}. Eq.~(\ref{eq:kink0}) fixes the kink angle that develops due to the presence of shear perturbations, while Eq.~(\ref{eq:kink1}) determines the subsequent curvature of the crack path. 

In order to use Eq.~(\ref{eq:kink0}), one needs to know the SIFs $K_\ell (\{h\},x)$ just before kinking for an arbitrary broken surface $h(x')$ (with $x'\le x$). For pure opening loading conditions and using a perturbation analysis around a straight crack (i.e. a crack parallel to the $x$-axis whose tip coincides with the curved crack located at $y=h(x)$), it can be shown that to first order in $h$ one can write $K_1 (\{h\},x)$ and $K_2 (\{h\},x)$ as functionals of $h(x)$ through~\cite{MovchanGao}
\begin{eqnarray}
K_1 (\{h\},x) &=& K^{*}_1(x)\,,\\
K_2 (\{h\},x)&=&K^{*}_2(x) + \frac{1}{2}h'(x)K^{*}_1(x) - \sqrt{\frac{2}{\pi}} \int_{-\infty}^x \frac{1}{\sqrt{(x-x')}}\frac{\partial}{\partial x'}\left\{\left(h(x')-h(x) \right) \sigma_{xx}^{*}(x')  \right\} dx' \label{eq:K2}\, ,
\end{eqnarray}
where $K^{*}_l$ and $\sigma_{xx}^{*}$ are the stress intensity factors and the $T$-stress component of a straight crack located at $y=h(x)$. Also, under pure tensile loading one readily has~\cite{MovchanGao}
\begin{eqnarray}
K^{*}_1(x)&=&K^{0}_1(x)\,,\\
K^{*}_2 (x) &\propto& h(x) K^{0}_1(x) + O\left(h^3\right)
\label{eq:K2star}.
\end{eqnarray}
Note that $K^{*}_2 (x)$ is proportional to $K^{0}_1 (x)$, as expected, but also depends on the geometry of the problem via $h(x)$. The superscript $0$ refers to quantities corresponding to the configuration of a centered straight crack (i.e. one that is located at $y=0$). In addition, we will assume, as in~\cite{Cotterell80}, that $\sigma_{xx}^{*}\equiv T$ is independent of $x$, arguing that its variation in space does not modify qualitatively the results. However, a variation in the stress intensity factor $K^0_1$ should induce a variation in the $T$-stress. Indeed, LEFM insures that for the same conditions under which Eq.~(\ref{eq:K2}) is valid, one has $\delta T/T= \delta K^0_1/K^0_1$. Notice that Eq.~(\ref{eq:K2star}) reveals an additional source of bias in the stability analysis of Cotterell and Rice, since the linear perturbation performed in~\cite{Cotterell80} ignores the term proportional to $h(x)$ in the expression of $K_2$.

At this point we introduce the heterogeneities in our model. The source of heterogeneities can be either variations of the elastic moduli in the material, or from residual stresses that were introduced for example by welding~\cite{Sumi92}, or during the machining of the material. Since the stress field is tensorial, these heterogeneities should affect both $K_1$ and $K_2$ independently. The local fluctuations in the toughness denoted by $k_1(x)$ have a finite mean $\overline {k_1}$. However, the local shear fluctuations, denoted by $k_2(x)$ must have a vanishing average because of PLS. Assuming that the crack advance between heterogeneities obeys the Principle of Local Symmetry and using Eqs.~(\ref{eq:kink0},\ref{eq:K2},\ref{eq:K2star}) and  the discussion above, one concludes that the local kinking angle is determined by
\begin{eqnarray}
\delta \theta (x) &=& -2 \delta \left(\frac {K_2}{K_{1}} \right) \simeq -2 \frac {\delta K_2}{K^0_{1}} + O\left( \delta K_\ell^2 \right) \nonumber \\
&\simeq& -\frac{2}{K_{1}}\left[ \delta k_2 + \left( \frac{1}{2}h'(x)+\frac{h(x)}{2H}-\sqrt{\frac{2}{\pi}}\frac{T}{\overline
{k_1}} \int_{-\infty}^x \frac{h'(x')}{\sqrt{x-x'}} dx' \right)\delta k_1 \right] + O\left( \delta K_\ell^2, \delta k_\ell^2 \right)
\label{eq:Dtheta},
\end{eqnarray}
where $H$ is a length-scale that depends on the geometry of the configuration. For example, it is proportional to the width of the strip in the case of a finite strip geometry - a configuration that is often adopted in experiments.

\begin{figure}[ht]
\centerline{\includegraphics[width=8cm]{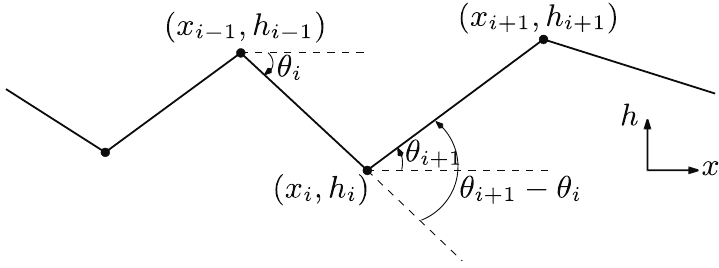}}
\caption{Schematic representation of the model - a crack path with straight kinks due to encounters with heterogeneities.}
\label{fig:model}
\end{figure}

The stage is set now to write the equation governing the crack path evolution. Since we are dealing with linear perturbations, we will also neglect the curvature parameter $a$ introduced in Eq.~(\ref{eq:shape}), and assume that the crack extension after kinking is always straight. This assumption is justified when $T=0$ or when the distances between  successive kinking events are small (which is equivalent to high density of heterogeneities). It leads to the configuration depicted in Fig.~\ref{fig:model}, from which one can easily read the equation $\theta_{i+1}-\theta_i=\delta \theta_i$. In the limit of small equal intervals $\delta x$ between successive heterogeneities, one has $\theta_i=h'_i$, then $h''_i=\delta \theta_i  / \delta x$ and Eq.~(\ref{eq:Dtheta}) leads to
\begin{equation}
h''(x)=-\frac{2}{K_{1}}\frac{\delta k_2}{\delta x} - \left(h'(x)+\frac{h(x)}{H}-2\sqrt{\frac{2}{\pi}}\frac{T}{\overline {k_1}}
\int_{-\infty}^x \frac{h'(x')}{\sqrt{x-x'}} dx'\right)\frac{1}{K_{1}}\frac{\delta k_1}{\delta x} \label{eq:motion}\,,
\end{equation}
where the indexes $i$ have been replaced by the position $x$ through the passage to the continuum limit. Eq.~(\ref{eq:motion}) reveals two noise terms that can be redefined by
\begin{eqnarray}
\eta_1'(x) &=& -\frac {1}{K_{1}} \frac{\delta k_1}{\delta x} \ , \\
\eta_2'(x) &=& -\frac {2}{K_{1}} \frac{\delta k_2}{\delta x}
\label{eq:etas1}.
\end{eqnarray}
Also, let us use the geometrical scale of the configuration $H$ as a unit length and define the constant
\begin{equation}
\beta \equiv 2\sqrt{\frac{2H}{\pi}}\frac{T}{\overline {k_1}}\,.
\end{equation}
Then, Eq.~(\ref{eq:motion}) becomes the following dimensionless stochastic equation
\begin{equation}
h''(x)=\eta_2'(x) + \eta_1'(x)\left( h'(x)+ h(x)-\beta \int_{-\infty}^x \frac{h'(x')}{\sqrt{x-x'}} dx' \right)
\label{eq:start} \ .
\end{equation}
Note that by choosing the length scale $H$, the total extension of the crack is not given. Also, $\beta$ can be either positive or negative depending on the sign of the $T$-stress. The discontinuous nature of crack propagation in a disordered material imposes a detailed discrete microscopic description of the influence of heterogeneities. The resulting stochastic integro-differential equation of the crack path should be derived as the continuum limit of the discrete model. This approach is different from previous pure continuum modeling~\cite{Ramanathan97b} that implicitly assumes smoothness of the paths and {\it one} source of noise that is introduced {\it a posteriori}. In opposite, Eq.~(\ref{eq:start}) shows that our approach leads to derivatives of {\it two} noise terms, one of which is multiplicative and the other is additive, without imposing them {\it a priori}.

The properties of the noise terms are prescribed by the original distribution of heterogeneities in the material that may exhibit long-range correlations as well as anisotropy. Although such features may be important~\cite{Mehdi} and in order to remain general, we assume short range correlations and thus model the noise terms $\eta_\ell(x)$ as independent Gaussian white noises
\begin{equation}
\left \langle \eta_\ell(x) \eta_m(x') \right \rangle = D_\ell
\delta_{\ell m} \delta (x-x') \label{eq:etas2}.
\end{equation}
Note that $\eta_1'(x)$ and $\eta_2'(x)$ that enter Eq.~(\ref{eq:start}) are conserved random terms (i.e., derivatives of white noises) modeling the fluctuations in the local toughness and the local shear respectively. In the following, we will show that the simple scenario of uncorrelated disorder already offers a rich spectrum of results. Including additional features in the disorder, such as long-range power law correlations, could lead to richer phenomena~\cite{Mehdi} and is left as a possible extension to the present analysis.

To be consistent with the derivation of the model one needs both the noise amplitudes $D_1$ and $D_2$ to be small. However, from Eq.~(\ref{eq:start}) one can see that varying the amplitude $D_2$ is equivalent to multiplying $h(x)$ by a constant, i.e. to fixing the overall scale of the height fluctuations, which does not influence the roughness of the curve. Since the scaling properties are not affected, the value of $D_2$ will not be reported in the following and $D_1$ will be the only pertinent  noise parameter. As a result $h(x)$ will be presented in arbitrary units. Regarding the $T$-stress, one expects $|\beta|$ to be of order $1$ in the framework of LEFM.

As a first observation, if the material is homogeneous or weakly disordered, one has $\eta_\ell'(x)\approx 0$ and the solution of Eq.~(\ref{eq:start}) is simply $h''(x)=0$. The addition of suitable initial conditions allows recovering the zero order solution $h(x)=0$ corresponding to a centered straight crack path. Eq.~(\ref{eq:start}) should be understood as resulting from a perturbation analysis of the crack trajectory around the solution in the absence of heterogeneities that is selected by the PLS. This should be contrasted with the stability analysis of a straight crack in a homogeneous material with respect to other solutions that satisfy also the PLS. An example of such a situation is the thermal crack problem~\cite{Yuse93,Marder94,adda95,thermique,Menouillard} where oscillatory crack paths exist in addition to the centered straight one and become more stable than the straight configuration at a given well defined threshold. The crack propagation there is always smooth and is very different from the stability encountered in disordered materials, which is due to the large density of heterogeneities that induces discontinuous propagation via linear segments between the heterogeneities.

\subsection{Numerical implementation}

In order to study crack paths that result from Eq.~(\ref{eq:start}), we start with a numerical integration of it. The initial condition will be always chosen to be a straight semi-infinite crack, $h(x)=0$ for $x \le 0$. In a discretized form, Eq.~(\ref{eq:start}) becomes\begin{equation}
h_{i+1}-2h_i+h_{i-1}=\Delta(\eta_{2,i}-\eta_{2,i-1}) +\Delta
(\eta_{1,i}-\eta_{1,i-1})\left( \frac{h_i-h_{i-1}}{\Delta}+
h_i-\frac{\beta}{\sqrt{\Delta}} \sum_{j=1}^{i-1}
\frac{h_j-h_{j-1}}{\sqrt{i-j}} \right) \label{eq:start-dis} \ ,
\end{equation}
for $i \ge 1$, using $\Delta$ as the uniform distance between heterogeneities and $h_0=h_1=0$ as initial conditions. Eq.~(\ref{eq:start-dis}) is a discretized version of Eq.~(\ref{eq:start}) that corresponds to the It\^{o} prescription \cite{ito} and was chosen by the discrete manner by which Eq.~(\ref{eq:start}) was derived. Also, the quantities $\eta_{\ell,i}$ are taken as independent random numbers, equally-distributed in the segment $\sqrt{d_\ell}\left[-1/2,1/2\right]$, and thus with variance $D_\ell=d_\ell/12$. The averages of $\eta_{\ell,i}$ are not important since only derivatives of the noise terms appear in the model. In order to be consistent with length normalisation, we pick a $1D$ lattice of size $L$ with $N$ sites, $N$ being also the number of heterogeneities in the interval $[0,L]$. Thus $\rho \equiv N/L=1/\Delta$ corresponds to the density of heterogeneities and a small $\Delta$ probes the regime of highly disordered materials.

\begin{figure}[ht]
\centerline{\includegraphics[width=10cm]{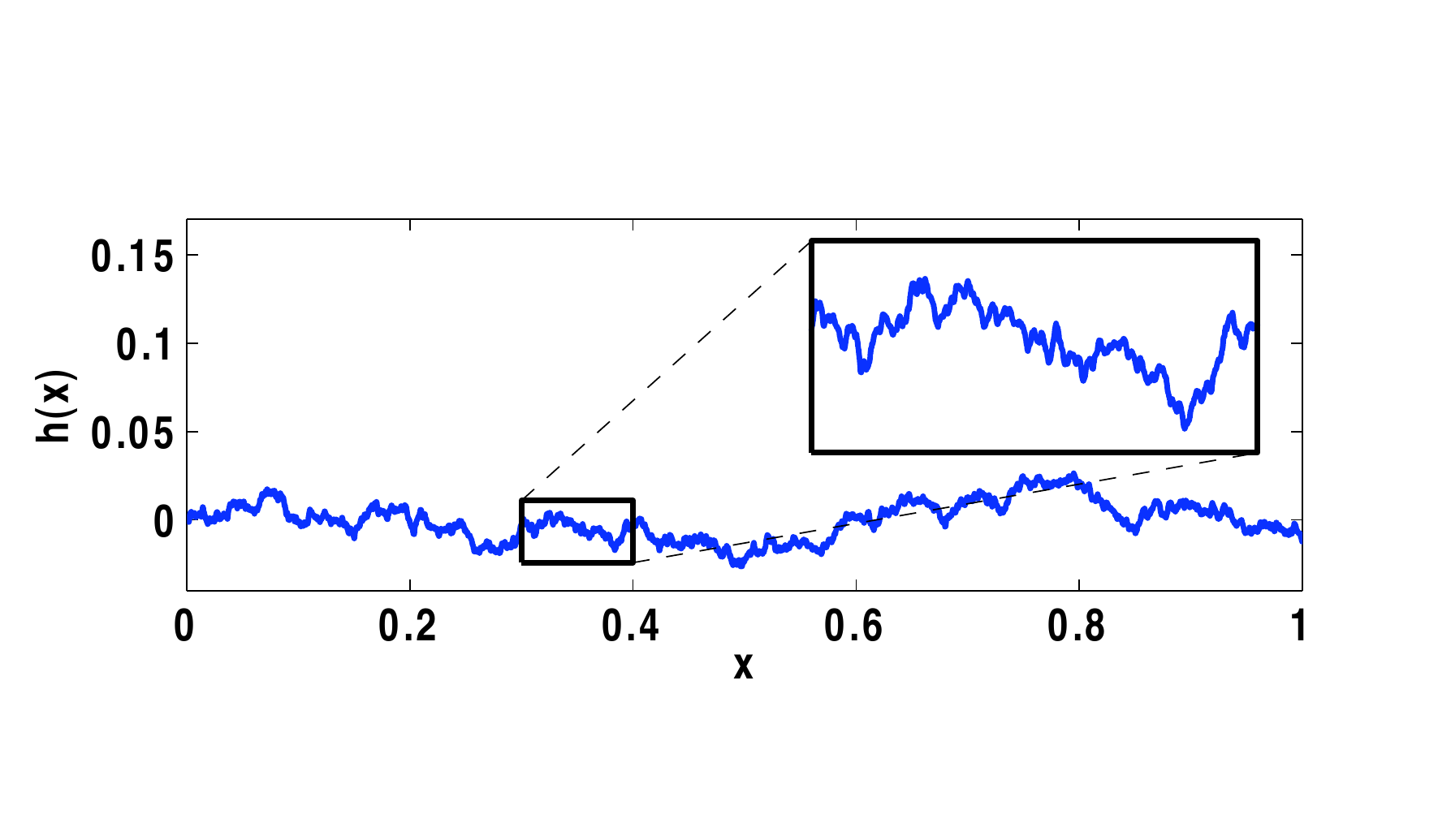}} 
\caption{An example of a crack path produced using the model, with $\Delta=10^{-5}$, $D_1=2\times 10^{-3}$ and $\beta=0$.
\label{fig:zoom}}
\end{figure}

Fig.~\ref{fig:zoom} shows an example of a crack grown using Eq.~(\ref{eq:start-dis}). The inset shows a zoom into a small part of the path, which may be suggestive of self-similar properties to the naked eye. However, before a thorough study of this aspect, a stability analysis \`a la Cotterell and Rice should be performed.

\subsection{Stability of crack propagation in disordered materials}

\begin{figure}[ht]
\centerline{\includegraphics[width=9cm]{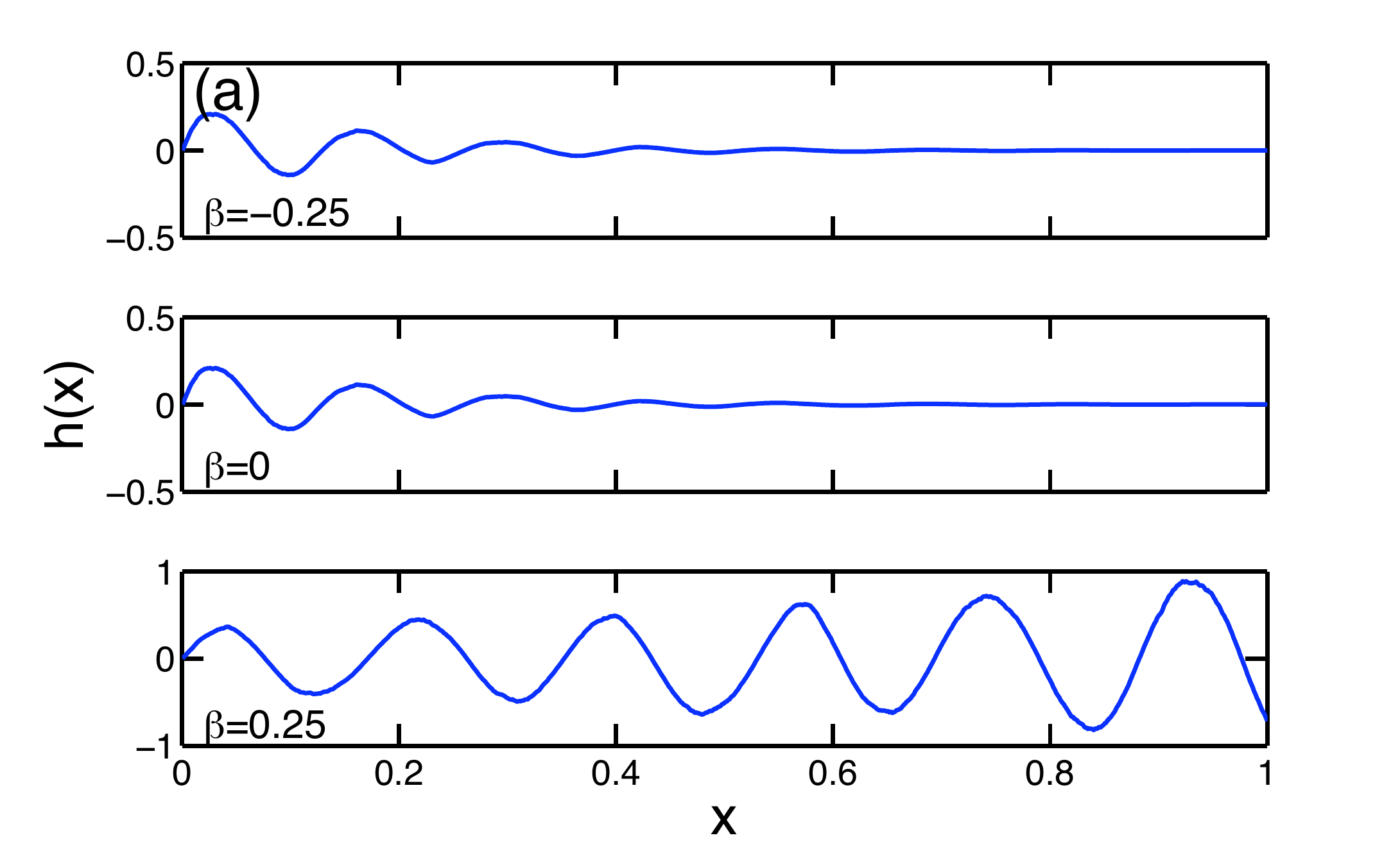}
\includegraphics[width=9cm]{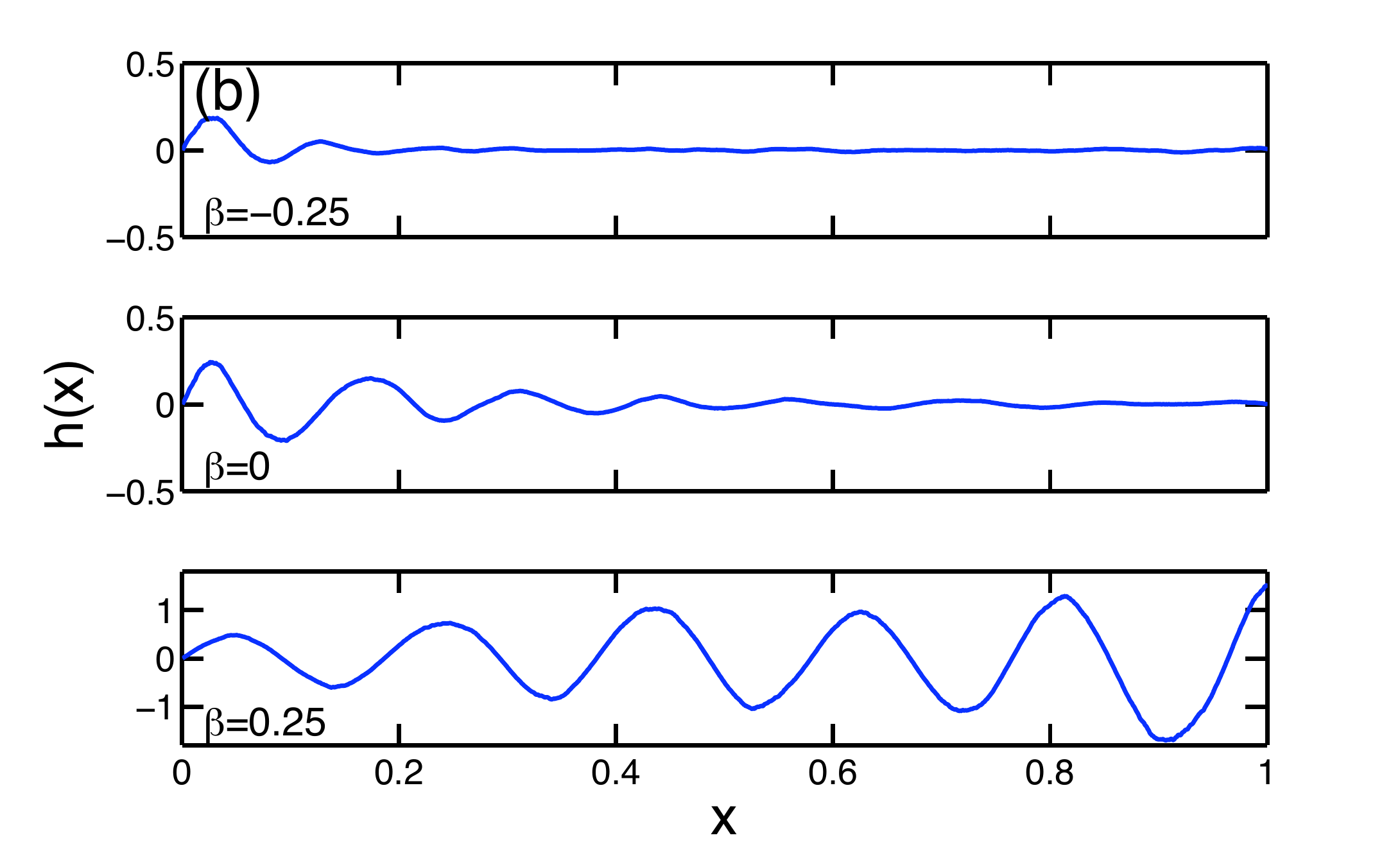}}
\caption{Examples of crack paths for various values of the $T$-stress and $D_2$ with $\Delta=0.1$, $D_1=0.1$. (a) $D_2=0$, i.e. $\eta_2(x)=\theta_0 \delta(x)$) and (b) $D_2=10^{-4}$. Note that for the cases $\beta=0.25$, the range of the $h$-axis is wider emphasising the exponential increase of the amplitude of the oscillations in those cases.
\label{fig:T-crit}}
\end{figure}

As mentioned above, the classical $T$-criterion of Cotterell and Rice~\cite{Cotterell80} states that straight tensile crack propagation in homogeneous materials become unstable when $T>0$. We therefore simulated crack paths with various values of $T$ in order to test this criterion within our model. We first consider the case with only one shear perturbation at $x=0$, after which the local toughness is $\eta_2=0$ and only $\eta_1$ is allowed to fluctuate. The results are presented in Fig.~\ref{fig:T-crit}a. Essentially, we recover the $T$-criterion, namely an instability occurs for $\beta>0$ (or equivalently $T>0$). It turns out that by adding the shear perturbations (i.e., $D_2 \ne 0$), the same scenario is recovered (see Fig.~\ref{fig:T-crit}b). One noticeable difference is that for positive values of $T$, the divergence of the path seems to accelerate due to the presence of the shear perturbations. Still, paths that do not experience shear perturbations ($T \le 0$ and $D_2=0$) will not destabilize in their presence. It should be mentioned that within the approach of Cotterell \& Rice it is not possible to follow more than one kink, as would certainly be the case in a disordered material where cracks propagate via many consecutive kinking events.

In summary, our results confirm the $T$-criterion for homogeneous materials and extends it to disordered systems. It is shown that straight crack propagation is unstable for $T>0$ and stable elsewhere. Moreover, in the stable case $T\le 0$ the marginal stability has been cured by the suppressing the square root behaviour $h \sim \sqrt{x}$ predicted in~\cite{Cotterell80}.

\section{Fractography}

In view of these results, from now on we will restrict our study to stable paths - as we are interested in crack roughness. We first focus on the case $T=0$, since it is simple enough to allow for definite numerical and analytical results, and at the same time contains the necessary complexity. This claim is based on scaling arguments - power counting~\cite{barabasi95} which is also supported by a numerical study. To put it simpler, as long as the crack path is stable, the presence of the $T$-stress does not change dramatically the shape of the crack paths. At the end of the paper, we will come back to this point and show how a non zero $T$-stress influences the result.

\subsection{The case $T=0$}

For the case of vanishing $T$-stress ($\beta=0$ in Eq.~(\ref{eq:start})), exact results have been obtained previously~\cite{EPL07}. Here, these results will be summarised briefly and extended. One technical difference is that in this work lengths are scaled using the width $H$ and not using the length $L$; as a result the dimensionless parameter $\alpha$ defined in~\cite{EPL07} is set to $1$ here. Essentially, our analysis is divided into two parts. The idea is to identify first the possible responses of the crack path to one shear perturbation, and only then to generalize to a superposition of many shear perturbations.

\subsubsection{Crack paths induced by a localized mode~II perturbation}

\begin{figure}[ht]
\centerline{\includegraphics[width=8cm]{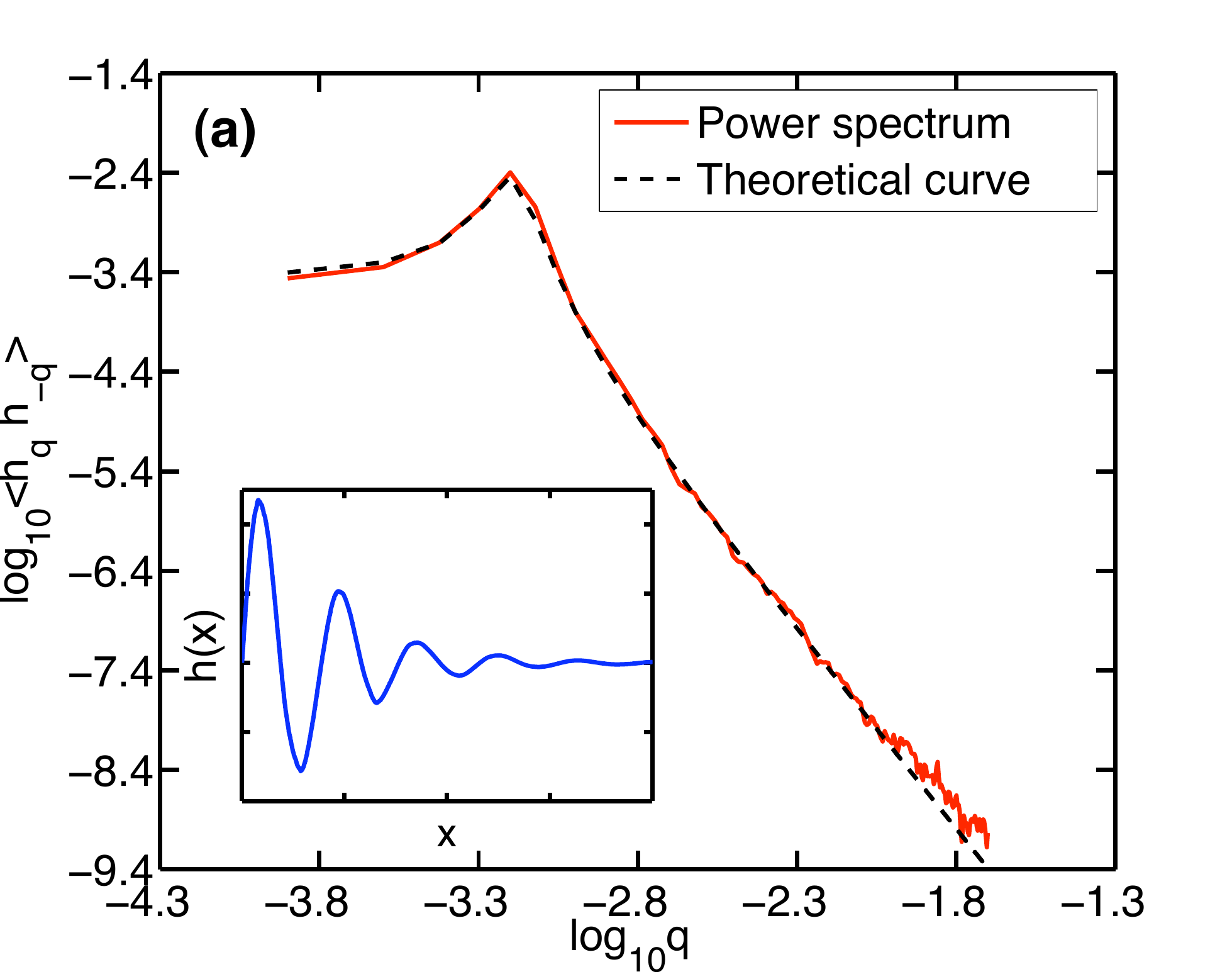} \includegraphics[width=8cm]{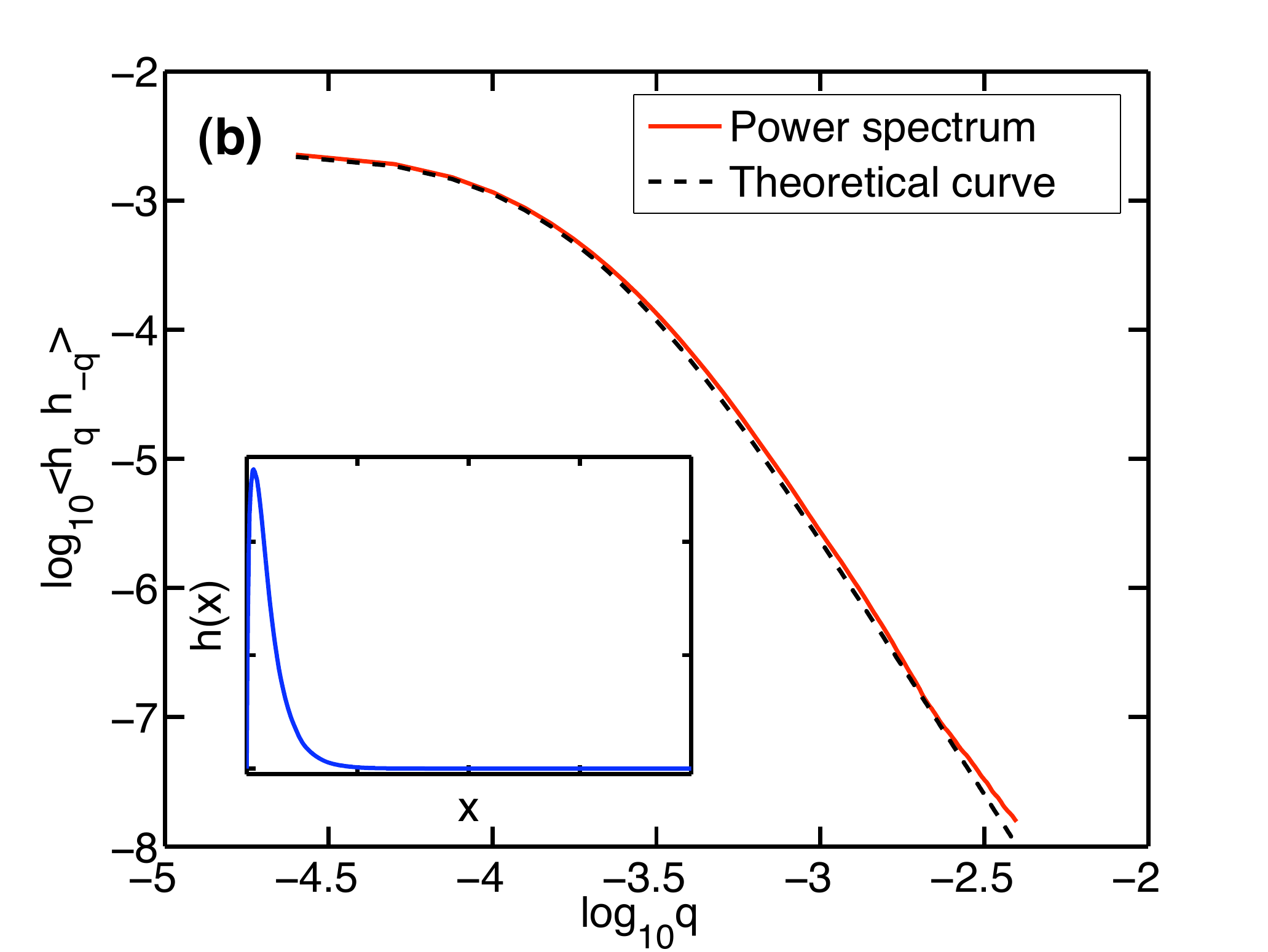}}
\caption{(Color online) Results of various cracks fractography for $\beta=0$ and $D_2=0$. The insets show examples of such paths, while the main figures show the power-spectra $\left\langle h_q h_{-q} \right\rangle_{\eta_1}$ averaged over $10$ realizations of the noise $\eta_1$.
Dashed curves are the corresponding theoretical curves. (a) The case $C<4$~: results produced using $\Delta=4\times10^{-4}$, $D_1=4\times10^{-5}$ with initial conditions $h(0)=0; h'(0)=1$.
(b) The case $C>4$~: results produced using $\Delta=4\times10^{-4}$, $D_1=2\times10^{-3}$, with initial conditions $h(0)=0; h'(0)=1$.
\label{fig:eta1}}
\end{figure}

Qualitatively, in reaction to a single shear perturbation, a straight crack deviates from its former direction by an angle which is proportional to the strength of the perturbation and then starts to relax to its original form. Interestingly, it is found that the crack path can relax in two different ways~: either by decaying exponentially (inset of Fig.~\ref{fig:eta1}a) or by decaying exponentially while
oscillating (inset of Fig.~\ref{fig:eta1}b).

This behaviour can be understood from the study of the logarithmic derivative of the crack path, $\psi(x)=h'(x)/h(x)$, which becomes stationary during the relaxation. A Fokker-Planck equation is derived for $\psi(x)$, along similar lines to those described in \cite{Derrida84}, and the effect of the toughness fluctuations, $\eta_1'(x)$, on average, can be reproduced by an effective deterministic evolution. This result allows to derive an effective (coarse grained) simple equation of motion in the presence of a single shear perturbation (and $\beta=0$), namely~\cite{EPL07}
\begin{equation}
 h''(x) = -C\left[ h'(x) + h(x) \right]
 \label{eq:h-c} \ ,
\end{equation}
where $C=D_1 / \Delta$. Comparing this with Eq.~(\ref{eq:start-dis}) one concludes that the averaged equation (\ref{eq:h-c}) is obtained from the full one (in a non-trivial way explained in \cite{EPL07}) by simply replacing the noise term $\eta_1'(x)$ with a constant, $-C$, that is always negative (even though $\eta_1'(x)$ is equally positive and negative), and proportional to its variance. This nontrivial result shows that the effect of the local toughness fluctuations is not so dramatic on the shape of the crack, apart from its constant variance and of course apart from setting a relevant scale for the energy that has to be invested in making the crack grow. However, one should be careful with the interpretation of Eq.~(\ref{eq:h-c}), as it describes only mean quantities, which does not imply that each realization behaves exactly the same.

Now, one can easily solve Eq.~(\ref{eq:h-c}) with the initial conditions $h(0)=0; h'(0)=\theta_0$ (i.e. one shear perturbation at the origin only). For $\left \langle h(x) \right \rangle_{\eta_1}$ we get
\begin{equation}
 \left \langle h(x) \right \rangle_{\eta_1} = \theta_0 e^{-\frac{1}{2} Cx} \frac{\sinh \left(\frac{1} {2}\sqrt {C\left( {C - 4 } \right)} x \right)}{\frac{1}{2} \sqrt{C(C - 4)}}
 \label{eq:h-csol} \, .
\end{equation}
This result shows why two kinds of responses to an initial perturbation are possible. The solutions of Eq.~(\ref{eq:h-c}) can exhibit either an exponential decay or damped oscillations depending on the sign of $C-4$. When $C > 4$, $h(x)$ simply decays exponentially, while for $C < 4$, the hyperbolic sine becomes an oscillating function, and thus we find an oscillatory relaxation. Since traditionally, noisy data are analysed in Fourier space by looking for example at the power spectrum, it would be interesting to obtain an analytical expression for it as well
\begin{equation}
 \left \langle h_q h_{-q} \right \rangle_{\eta_1} = \frac{\theta_0^2} {\left(q^2 - C\right)^2 + C^2 q^2}
 \label{eq:h-cPS} \, ,
\end{equation}
where $h_q$ is the Fourier component of $h(x)$. In Figs.~\ref{fig:eta1}a-\ref{fig:eta1}b below we compare the result of the averaged power spectrum over $10$ simulated paths (all with the same parameters but different realizations of the noise) for the two cases $C<4$ (damped oscillations) and $C>4$ (i.e., simple exponential decay). As can be seen the theoretical curve is in very good agreement with the numerical result over many decades.

\subsubsection{Crack paths in the presence of an extended mode~II perturbation}

\begin{figure}[ht]
\centerline{\includegraphics[width=8cm]{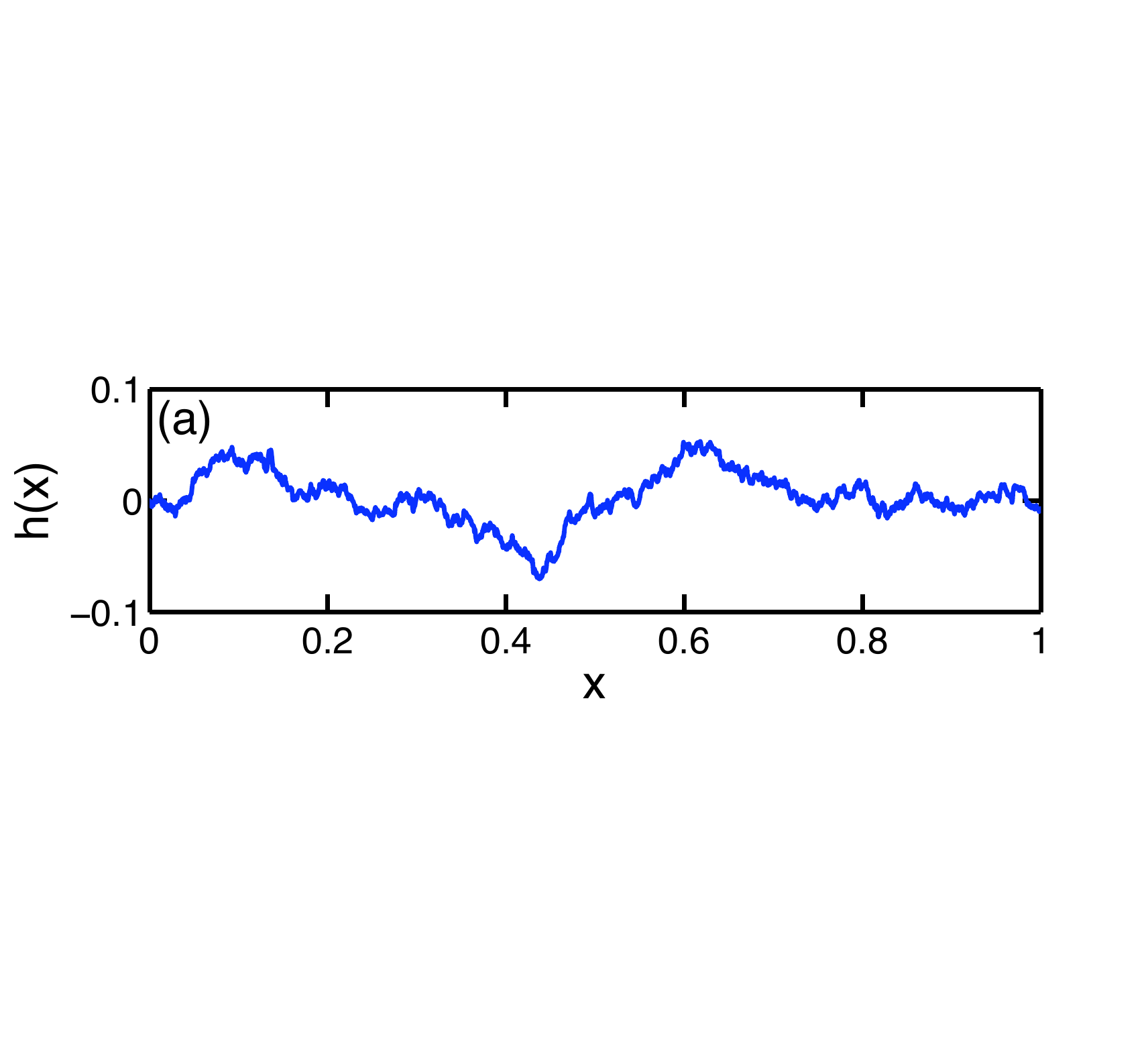}}
\centerline{\includegraphics[width=8cm]{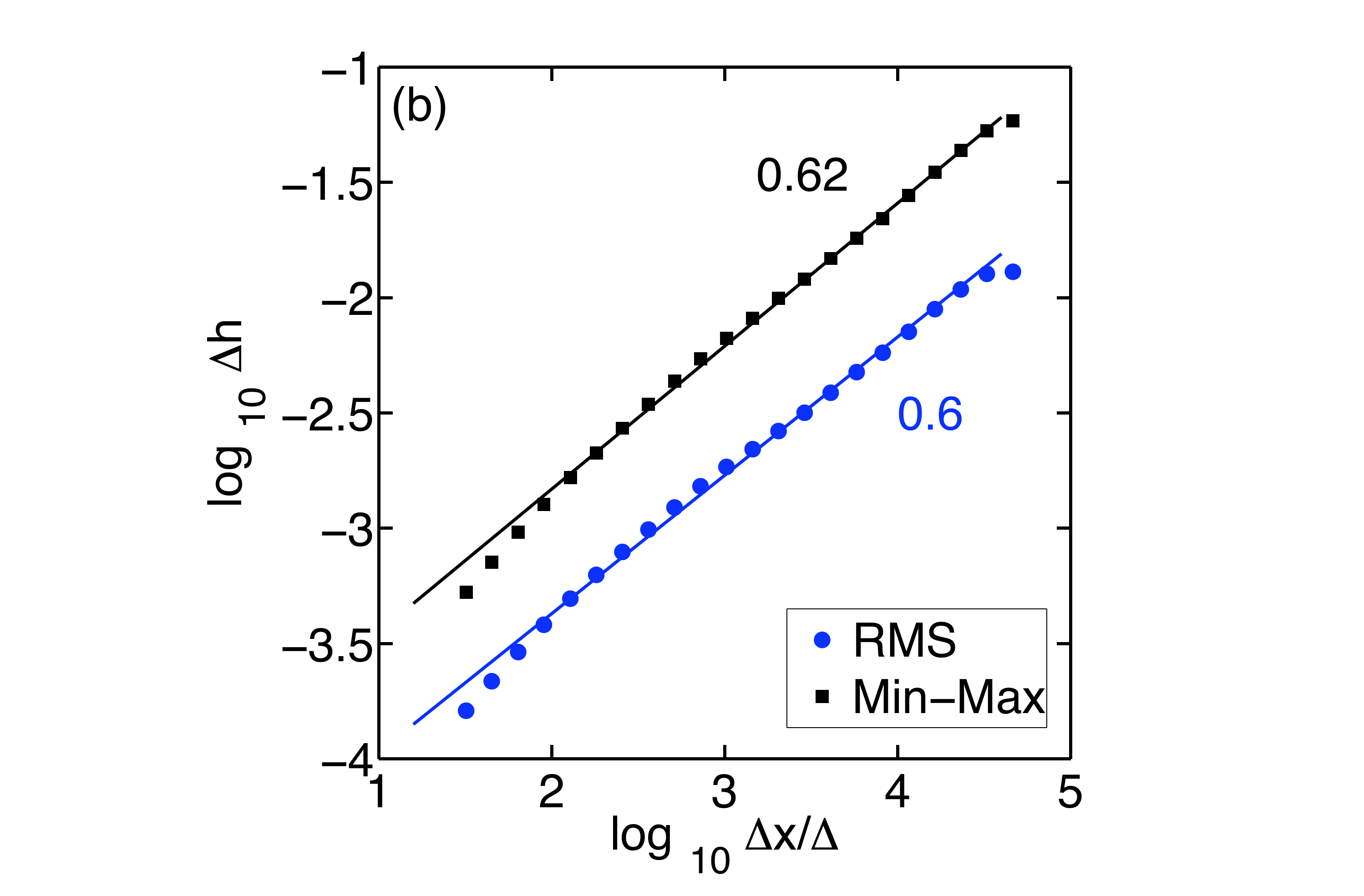}
\includegraphics[width=8cm]{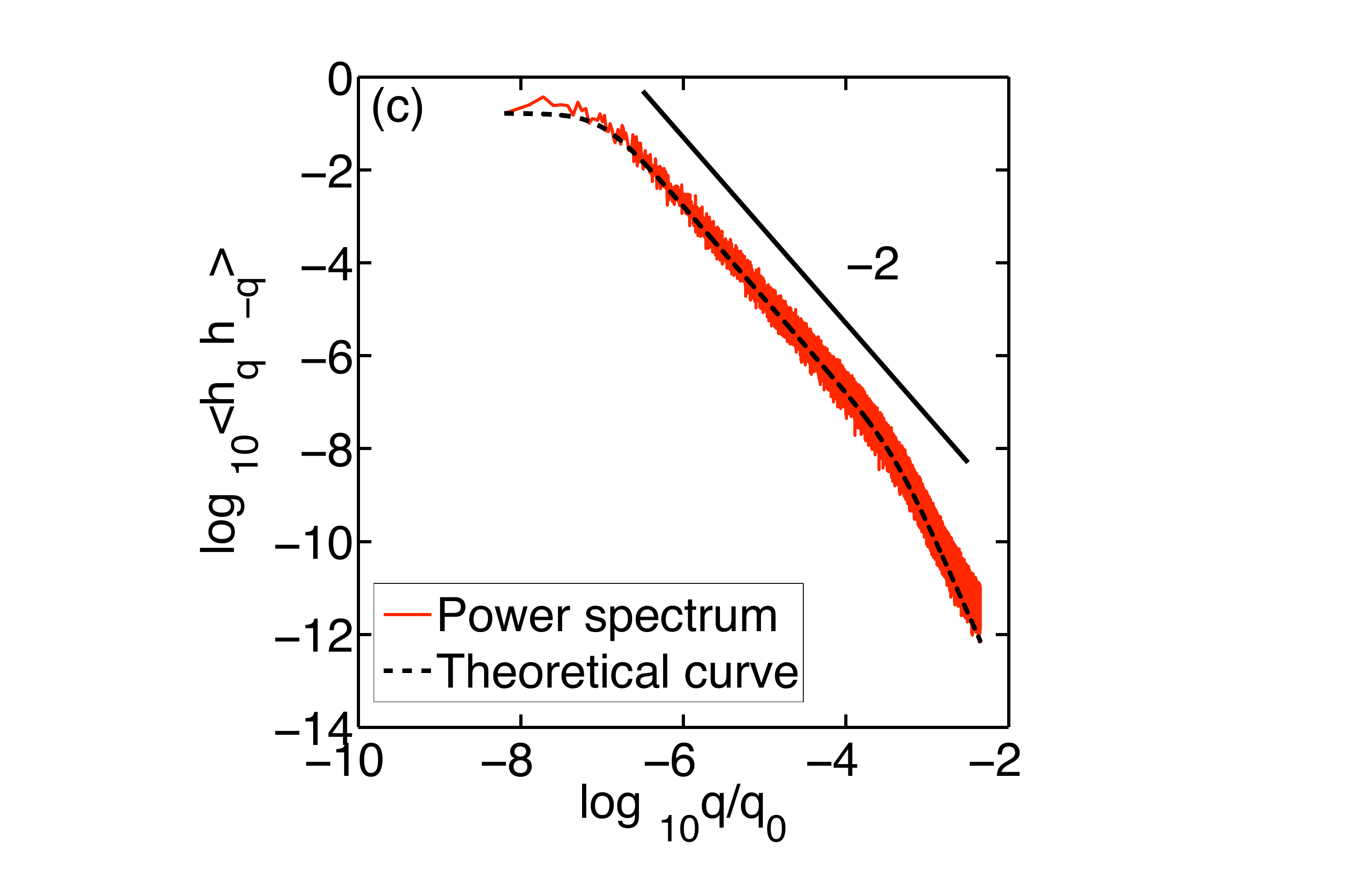}}
\caption{(Color online) (a) A crack path simulated using $\Delta=10^{-4}$ and $D_1=4\times10^{-2}$. (b) Results of a Min-Max and RMS analysis that yields $\zeta\simeq 0.6$ over more than $2$ decades. (c) The power spectrum of the crack path and a theoretical prediction for it based on the parameters of the model averaged over $10$ realizations.
\label{fig:0.6}}
\end{figure}

A natural step forward is to study crack propagation in a regime where there are many shear perturbations. This of course amounts to retaining the additive noise $\eta_2'(x)$ in Eq.~(\ref{eq:start}). Unlike the fluctuations in the local toughness, the shear perturbations cannot be modeled by a constant. This term seems crucial for creating the random patterns that are observed for fracture surfaces in nature. Interestingly, varying the various parameters results in rather different patterns, as shown in Fig.~\ref{fig:0.6}a-\ref{fig:0.8}a. Moreover, when analyzed using the real space methods, such as the Min-Max or the RMS method (see Appendix), one can produce various values of roughness exponent $\zeta$ which depend on the parameters of the model. Figs.~\ref{fig:0.6}b-\ref{fig:0.8}b show the analysis the two crack paths and show that one can obtain, for example, values of roughness exponent  $\zeta \simeq 0.6$ and $\zeta  \simeq 0.8$ that can be found in literature~\cite{Plouraboue96,Balankin05,Salminen03,Proc06,Mallick06,Wood,Straws}.

\begin{figure}[ht]
\centerline{\includegraphics[width=8cm]{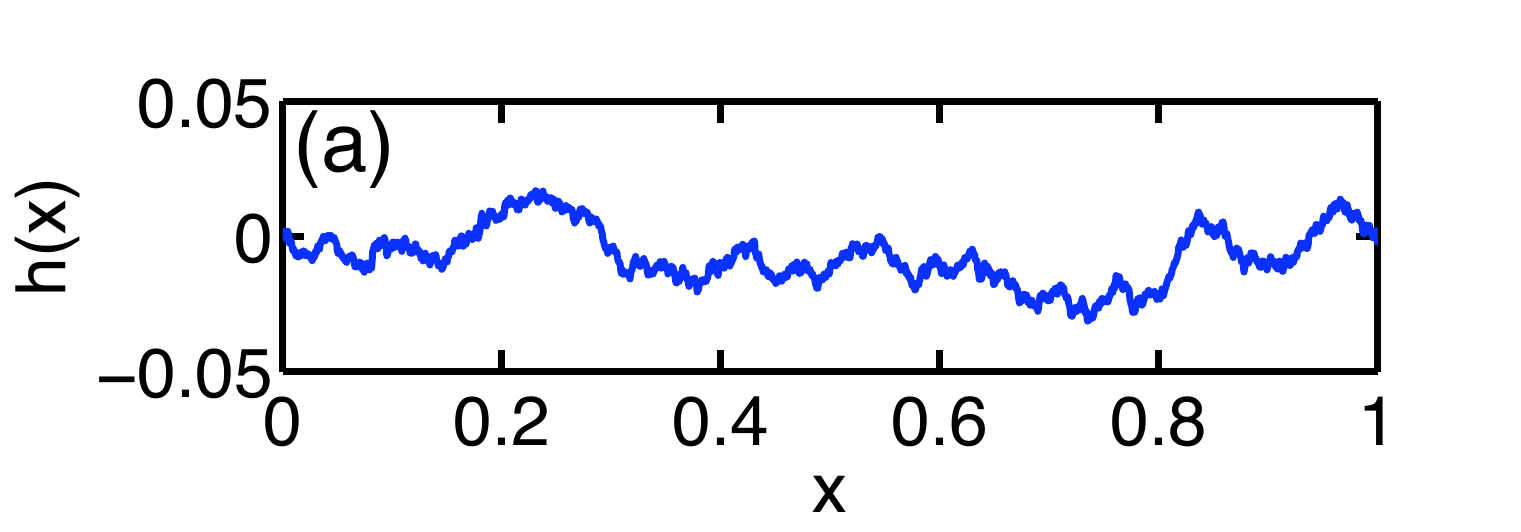}}
\centerline{\includegraphics[width=8cm]{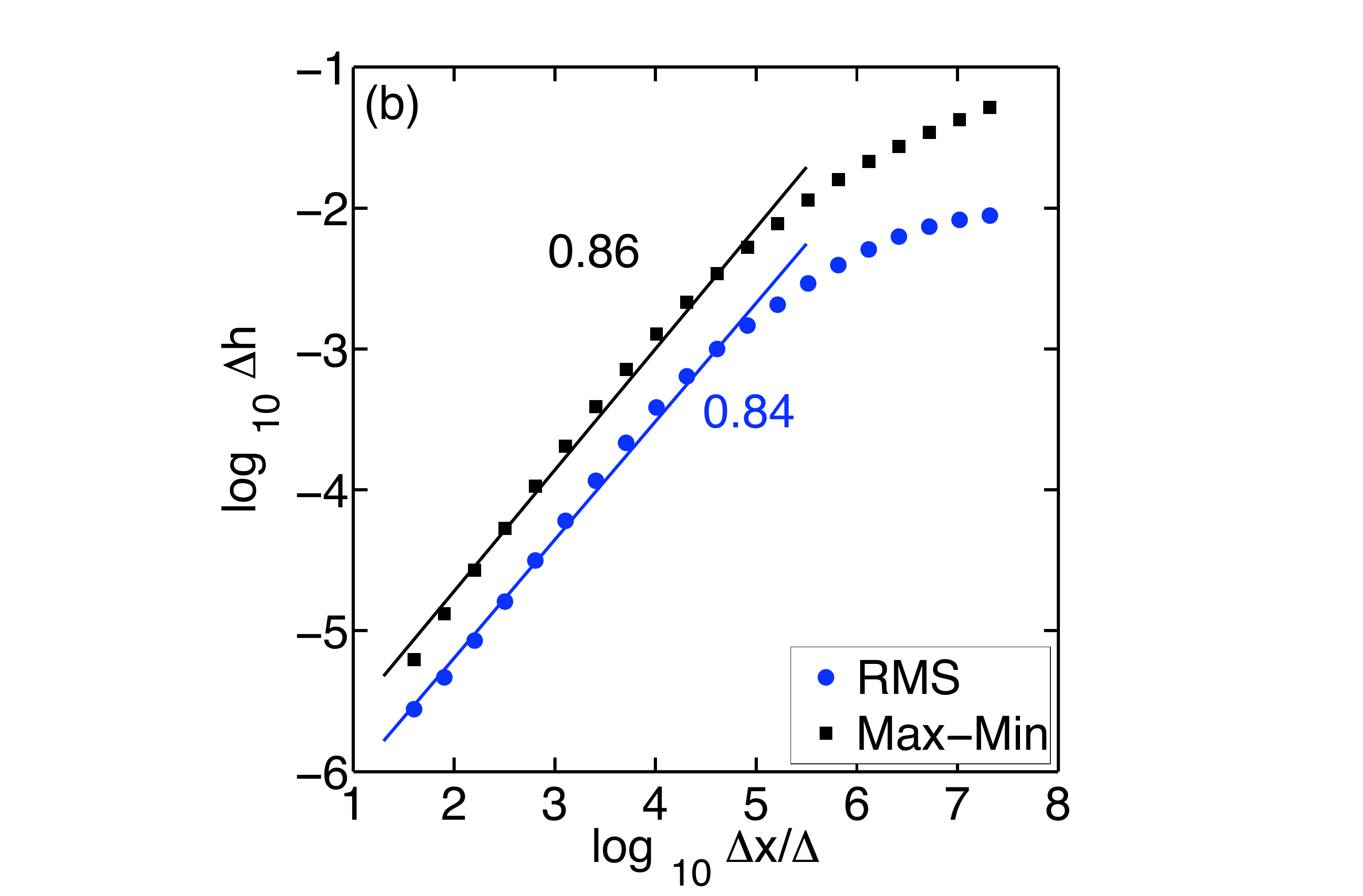}
\includegraphics[width=8cm]{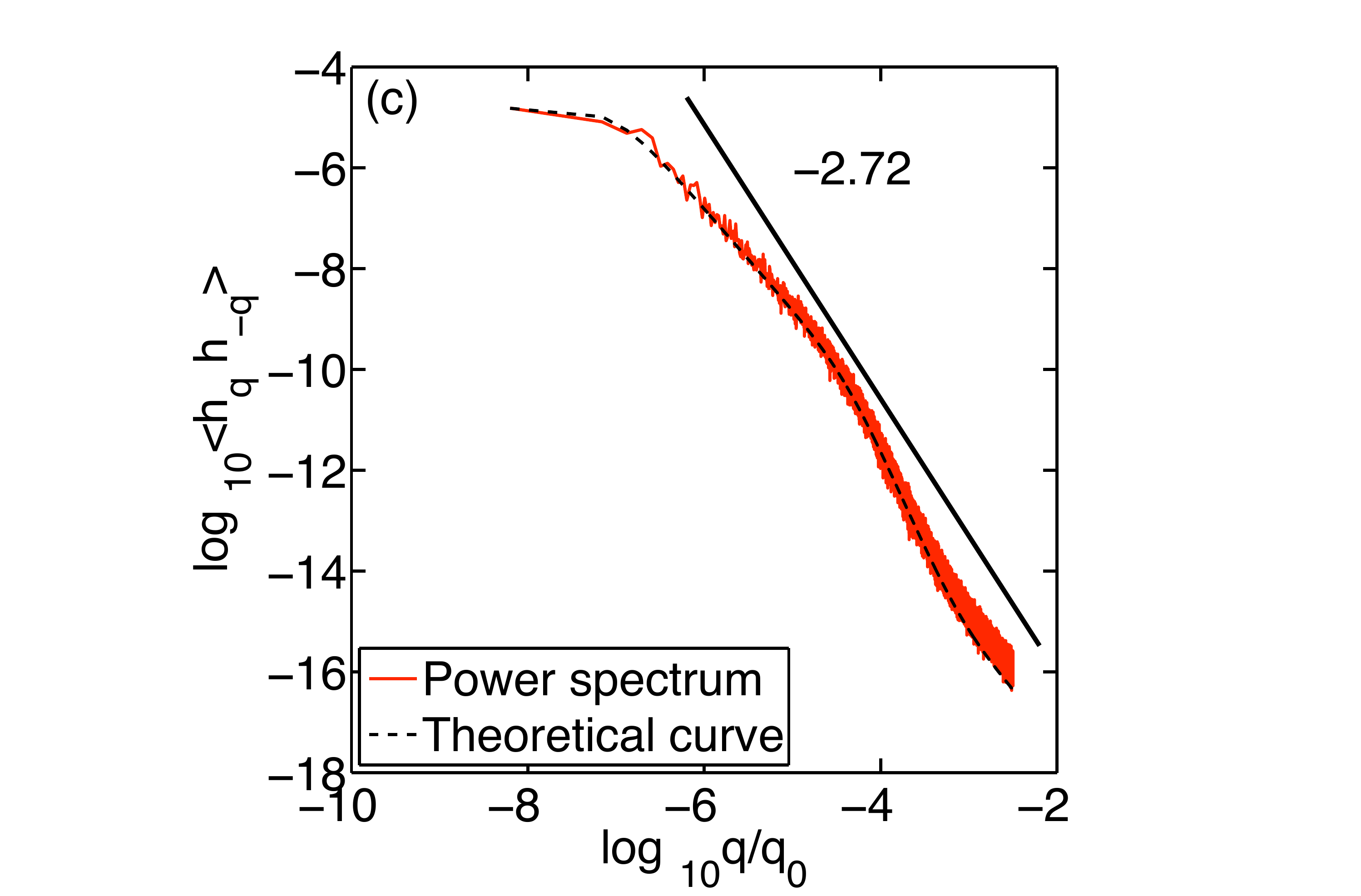}}
\caption{(Color online) (a) A crack path simulated using $\Delta=10^{-6}$ and $D_1=4\times10^{-4}$. (b) Results of a Min-Max and RMS analysis that yields $\zeta\simeq 0.8$ over a decade and a half. (c) The power spectrum of the crack path and a theoretical prediction for it based on the parameters of the model averaged over $10$ realizations.
\label{fig:0.8}}
\end{figure}

However, when looking at the power-spectrum of each crack path there is no longer a simple scaling picture, as in the Min-Max or RMS plots. For both cracks, the power spectrum always begins with a plateau for small values of $q$, but for larger values of $q$ there isn't an easy way to determine a slope, which seems to vary over different values. This phenomenon is traditionally interpreted as a crossover between different regimes characterized by different roughness exponents in the analysis of experimental data~\cite{Bonamy06}. In view of the analytical results which we obtained in the previous section we argue for a different and simpler scenario.

The starting point is Eq.~(\ref{eq:start}) with $\beta=0$, supplemented with the initial condition $h(0)=0$ and $h'(0)=0$. In this equation $\eta_2'(x)$ is just a nonhomogeneous term. This means that once we have a solution for the homogeneous equation, we can build a special solution that solves the nonhomogeneous part. Recalling that $\eta_1(x)$ and $\eta_2(x)$ are independent random variables, we conclude that, as before, averaging over realizations of the local toughness fluctuations amounts to replacing $\eta_1'(x)$ by $-C$. This yields
\begin{equation}
 h''(x) = -C\left[ h'(x) + h(x) \right] + \eta_2'(x)
 \label{eq:h2-c} \ .
\end{equation}
Using the initial conditions $h(0)=h'(0)=0$, the solution of this equation is given by
\begin{equation}
h(x) = \int\limits_0^x {h_0 \left( {x - y} \right)\eta'_2 \left( y
\right)dy}
 \label{eq:h2-sol} \ ,
\end{equation}
where $h_0(x)$ is the solution to the homogeneous problem given by
\begin{equation}
 h_0(x) = e^{-\frac{1}{2}Cx} \frac{\sinh \left({\frac{1}{2}\sqrt{C(C - 4)}x}\right)}{\frac{1}{2}\sqrt{C(C - 4)} }
 \label{eq:h0-sol} \ .
\end{equation}
The power spectrum of this solution yields the expression~\cite{EPL07}
\begin{equation}
 \left \langle h_q h_{-q} \right \rangle_{\eta_1,\eta_2}  = D_2 \frac{\left(1+C^{-1}\right) q^2 + (2C+3)} {\left(q^2 - C \right)^2 + C^2 q^2}
 \label{eq:h2-cPS} \ .
\end{equation}
This expression is different from the expression given by Eq.~(\ref{eq:h-cPS}) as there is a $q^2$ term in numerator, which implies a tail of $1/q^2$ in the spectrum. The coefficients in Eq.~(\ref{eq:h2-cPS}) are determined from the Fourier transform of a stationary signal after cutting out the transient regime, that is $h_q=\int_{x_0}^1 h(x) e^{iqx}dx$. This leads to a system with effective random initial conditions at $x=x_0$. Since $x_0$ is chosen in the steady regime the statistics of $h(x_0)$ and $h'(x_0)$ are known, and an analytical expression of the power spectrum can be obtained. Figs.~\ref{fig:0.6}c-\ref{fig:0.8}c shows that despite Eq.~(\ref{eq:h2-cPS}) agrees very well with the simulations data for the power spectrum, one might be tempted to fit it with a power law ansatz. However, apart from a tail $q^{-2}$ at large $q$'s, which yields a roughness exponent $\zeta=0.5$ at small length scales, Eq.~(\ref{eq:h2-cPS}) tells us that there is no self-affine behaviour of fracture surfaces at intermediate length scales, a simple crossover is taking place.

The result contained in Eq.~(\ref{eq:h2-sol}) allows us to derive the full Probability Distribution Function of $\Delta h (\Delta x)$ as defined in Eq.~(\ref{eq:delh}), which is becoming a popular measure for self-affinity~\cite{Santucci07,Auradou07,Horst09,Bonamy09}. Using Eq.~(\ref{eq:h2-sol}) and some simple manipulations one can rewrite $\Delta h (\Delta x)$ as
\begin{equation}
\Delta h(\Delta x) = \int_{-\infty }^0 {h'_0 (- x) \left[ {\eta_2
\left(x + \Delta x\right) -  \eta_2(x)} \right]dx}
 \label{eq:delh2} \, .
\end{equation}
Here, the left most point has been pushed to $-\infty$ in order to ensure stationarity. Then, the required PDF is formally given by
\begin{equation}
P\left( {\Delta h\left( {\Delta x} \right)} \right) = \left\langle
{\delta \left( {\Delta h\left( {\Delta x} \right) - \int_{ - \infty
}^0 {h'_0 \left( { - x} \right)\left[ {\eta _2 \left( {x + \Delta x}
\right) - \eta _2 \left( x \right)} \right]dx} } \right)}
\right\rangle_{\eta_2}
 \label{eq:PDF1} \, .
\end{equation}
Using the Fourier representation of the Delta distribution $\delta(x) = \frac{1}{2\pi }\int\limits_{-\infty}^\infty  {e^{iqx} dq}$, Eq.~(\ref{eq:PDF1}) becomes
\begin{equation}
P\left( {\Delta h} \right) = \frac{1}{{2\pi }}\int\limits_{ - \infty
}^\infty  {e^{iq\Delta h} \left\langle {e^{ - iq\int_{ - \infty }^0
{h'_0 \left( { - x} \right)\left[ {\eta _2 \left( {x + \Delta x}
\right) - \eta _2 \left( x \right)} \right]dx} } } \right\rangle
_{\eta _2 } dq}
\label{eq:PDF2} \, .
\end{equation}
Since the term in the exponent is a linear combination of independent random terms one gets
\begin{equation}
\left\langle {e^{ - iq\int_{ - \infty }^0 {h'_0 \left( { - x}
\right)\left[ {\eta _2 \left( {x + \Delta x} \right) - \eta _2
\left( x \right)} \right]dx} } } \right\rangle  = e^{-\frac{1}{2}q^2
\left\langle {\left[ {\int_{ - \infty }^0 {h'_0 \left( { - x}
\right)\left[ {\eta _2 \left( {x + \Delta x} \right) - \eta _2
\left( x \right)} \right]dx} } \right]^2 } \right\rangle }
\label{eq:PDF3} \, ,
\end{equation}
namely a Gaussian, and one just needs to calculate its variance
\begin{eqnarray}
&& \sigma ^2 \left(\Delta x\right) \equiv \left\langle \left[
{\int_{-\infty }^0 {h'_0(-x) \left[ {\eta _2 \left( {x + \Delta x}
\right) - \eta_2(x)} \right]dx}} \right]^2 \right\rangle_{\eta_2}  \nonumber \\
&&= \frac{D_2}{C}\left[1 - \exp\left\{ - C\Delta x
{\textstyle{{4 - C\cosh \left[ {\frac{1}{2}\sqrt {C(C - 4)} \Delta x} \right] + \sqrt {C(C - 4)} \sinh \left[ {\frac{1}{2}\sqrt {C(C - 4)} \Delta x} \right]} \over {C - 4}}} \right\} \right]
\label{eq:PDF4} \, ,
\end{eqnarray}
and finally
\begin{equation}
P\left( {\Delta h\left( {\Delta x} \right)} \right) = \frac{1}{\sqrt
{2\pi \sigma ^2 \left( {\Delta x} \right)}} e^{ - \frac{1}{2}\left[
{\frac{{\Delta h\left( {\Delta x} \right)}}{{\sigma \left( {\Delta
x} \right)}}} \right]^2 } \label{eq:PDF5} \, .
\end{equation}

Recall that for the underlying shape $h(x)$ to be self affine the PDF must obey two properties (see Eq.~(\ref{scaling4})): First, it should have the same form for all the scales $\Delta x$. This property is explicitly obeyed by the derived PDF (\ref{eq:PDF5}). The second requirement is that the RMS $\sigma(\Delta)$ scales as $(\Delta x)^{\zeta}$. This property is not obeyed here, since
\begin{equation}
\sigma \left(\Delta x\right) \sim\left\{ \begin{array}{l}
 \sqrt{2D_2 \Delta x} \quad \Delta x \ll 1 \\
 \sqrt{D_2/C} \quad \;\;\Delta x \gg 1 \\
 \end{array} \right.
\label{eq:PDF6} \, ,
\end{equation}
meaning that there is a slow crossover from a square-root behaviour, for small scales, to a constant, for large scales, and strictly speaking the path $h(x)$ is not self-affine.

\begin{figure}[ht]
\centerline{\includegraphics[width=8cm]{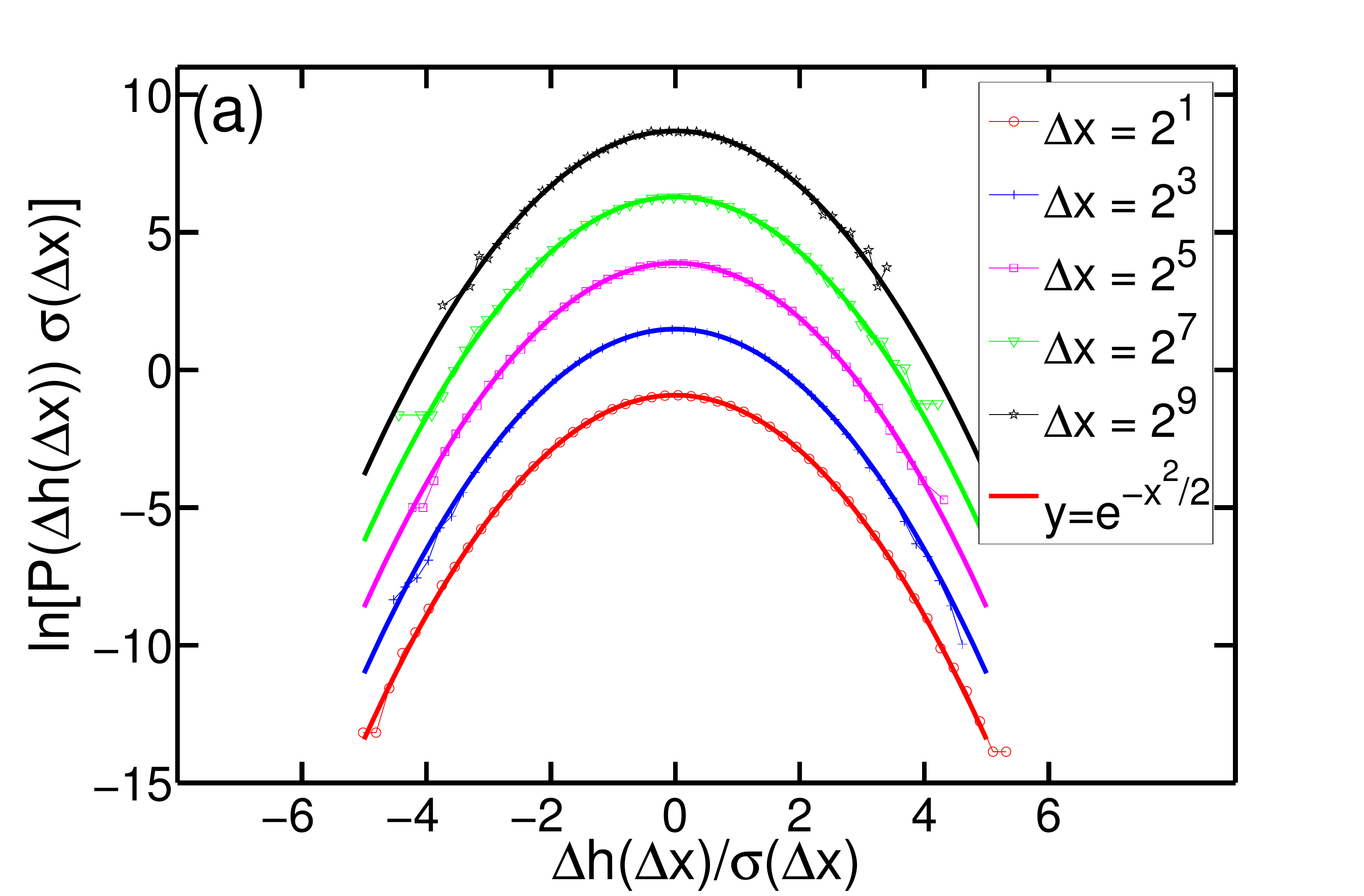} \includegraphics[width=8cm]{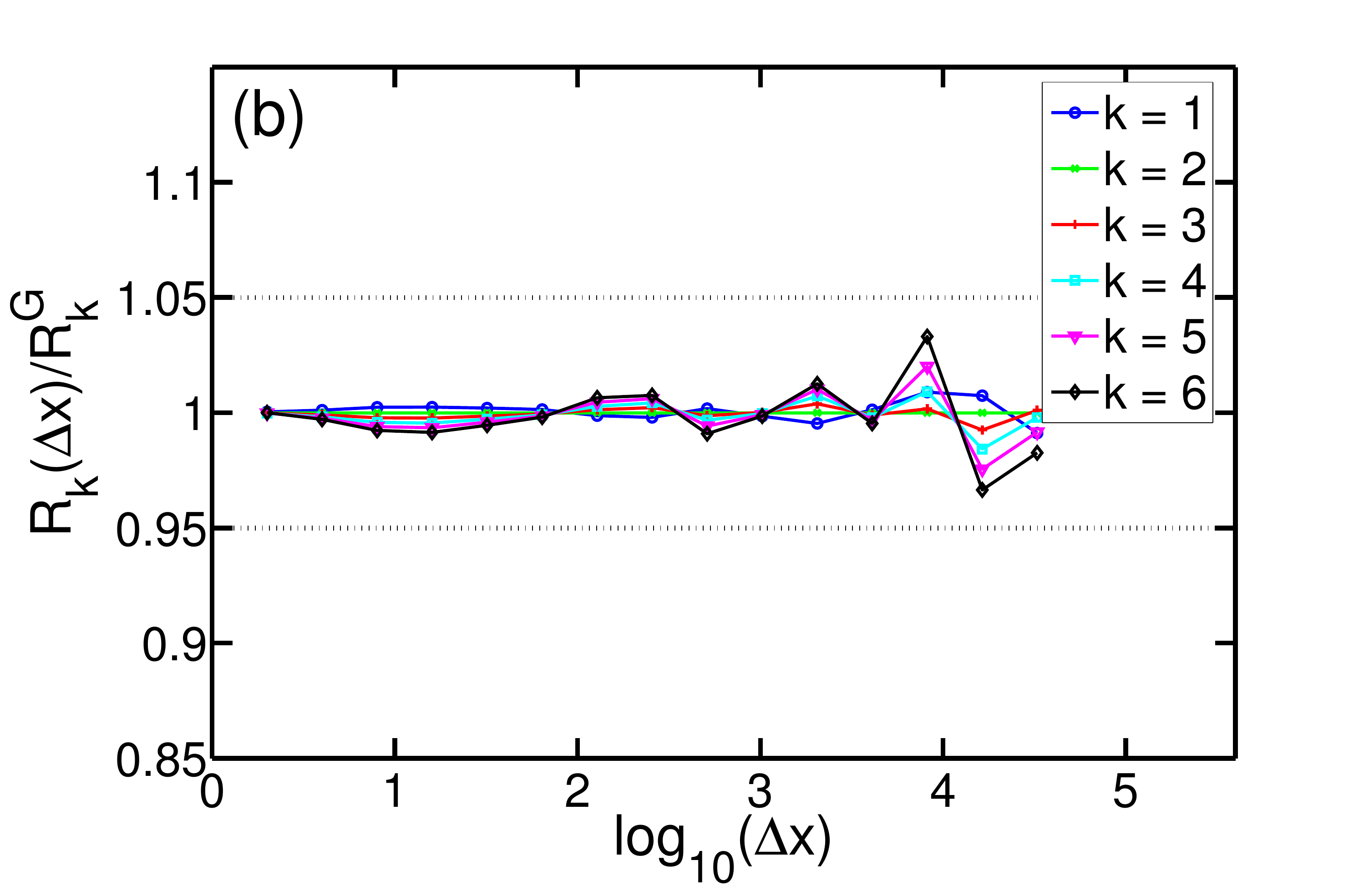}}
\caption{(Color online) (a) Statistical normalized distribution of height fluctuations $P(\Delta h (\Delta x) )$ sampled from a crack path simulated using $\Delta=10^{-4}$ and $D_1=4\times10^{-2}$ over $10^7$ time points on a semi-log scale. For each $\Delta x$ a pure
Gaussian is plotted as a guide to the eye. Note that the various distributions are shifted logarithmically horizontally for visual clarity. (b) A comparison of the moments $R_k(\Delta x)$ to those expected for a Gaussian distribution $R_k^G$ for $k=1,\ldots ,6$. The ratio $R_k(\Delta x)/R_k^G$ is presented and the dashed lines mark a $5\%$ deviation interval.
\label{fig:0.6PDF}}
\end{figure}
In order to check these theoretical predictions, a propagating crack has been simulated over a very long interval, using the parameters presented in Fig.~\ref{fig:0.6}, in order to produce the PDF for various values of $\Delta x$. As can be seen, the height distribution seems to exhibit Gaussian statistics as predicted. In order to verify this more quantitatively, we compare the moments of $\Delta h(\Delta x)$ normalized by the $2^{nd}$ moment, namely
\begin{equation}
R_k(\Delta x) \equiv \frac{\left \langle |\Delta h (\Delta x)|^k
\right \rangle^{1/k}} {\left \langle |\Delta h (\Delta x)|^2 \right
\rangle^{1/2}} \label{eq:Rk} \, ,
\end{equation}
to those obtained by the Gaussian distribution, namely
\begin{equation}
R_k^G = \sqrt{2} \left(\frac{\Gamma
\left(\frac{k+1}{2}\right)}{\sqrt{\pi}}\right)^{1/k} \label{eq:RkG}
\, ,
\end{equation}
up to $6^{th}$ order. The results are presented in Fig.~\ref{fig:0.6PDF}b, and supports the Gaussianity of the distributions.

\begin{figure}[ht]
\centerline{\includegraphics[width=8cm]{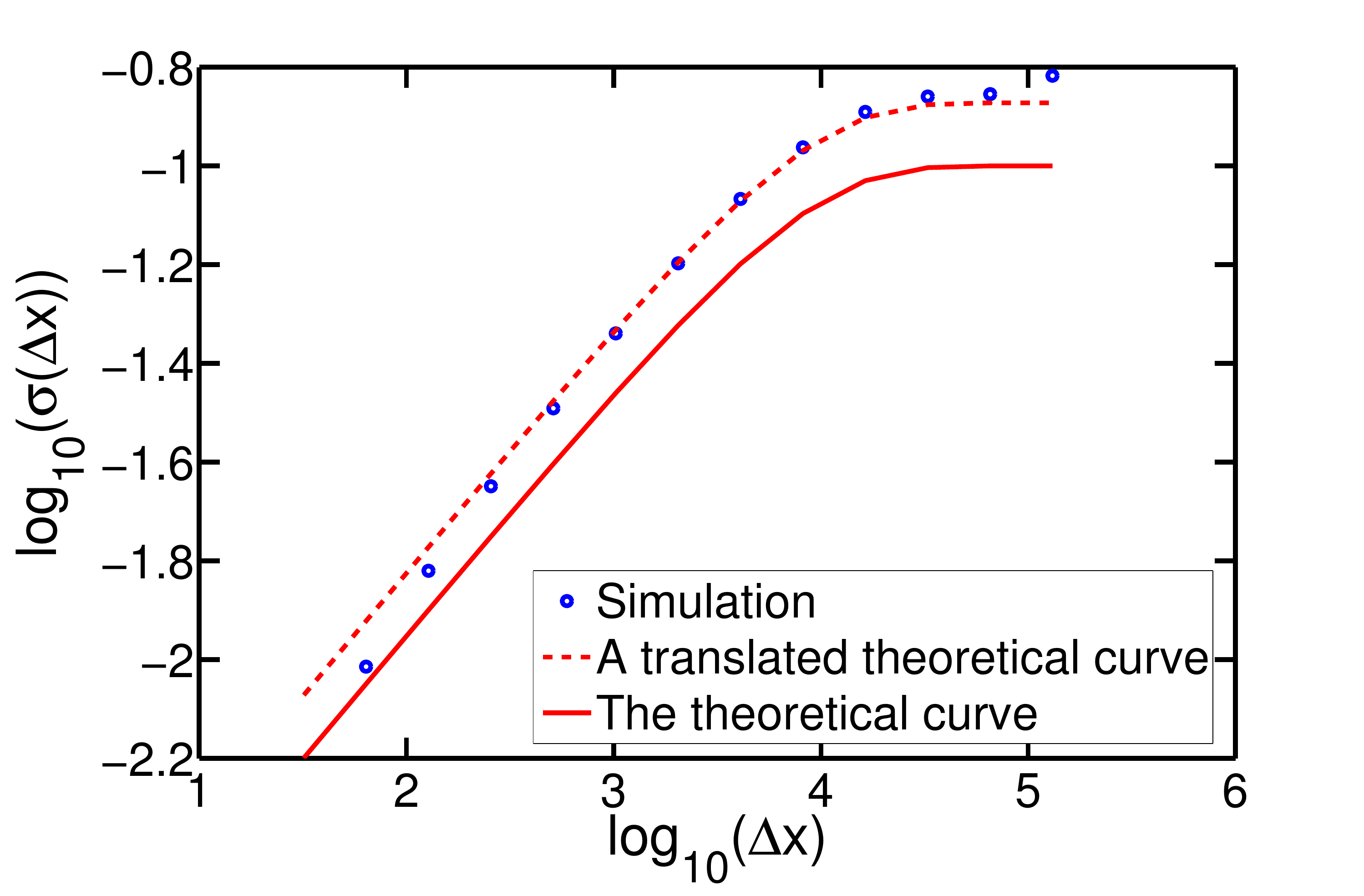}}
\caption{(Color online) A comparison of $\sigma(\Delta x)$ obtained from the simulation (with the parameters defined in Fig.~\ref{fig:0.6PDF}) to the one calculated in Eq.~(\ref{eq:PDF4}).
\label{fig:0.6PDFsigma}}
\end{figure}

Finally, we compare the width of distribution $\sigma(\Delta x)$ to the one given by Eq.~(\ref{eq:PDF4}). As can be seen in Fig.~\ref{fig:0.6PDFsigma}, although the theoretical prediction captures the form and has the right order of magnitude, it clearly deviates from the result of the simulation. In fact, it seems that the prefactor in Eq.~(\ref{eq:PDF4}) (i.e. $D_2/C$) underestimates the measured one, such that by tuning this prefactor one can reproduce the right behaviour over the whole range. A possible reason for this difference is due to discretization and finite-size scaling. Another reason could be the fact that we derive the PDF by first averaging over $\eta_1$ (and thus obtaining the effective equation (\ref{eq:h2-c})), and only then averaging over $\eta_2$, while in reality these two noisy terms fluctuate simultaneously at the same scale. Since the PDF is a sensitive probe this delicate issue is pronounced. At any rate, the simulated $\sigma(\Delta x)$ confirms the statement that there is a crossover from a square-root behaviour to a constant, and thus no real self affinity exists.

\subsection{The effect of the $T$-stress}

\begin{figure}[ht]
\centerline{\includegraphics[width=8cm]{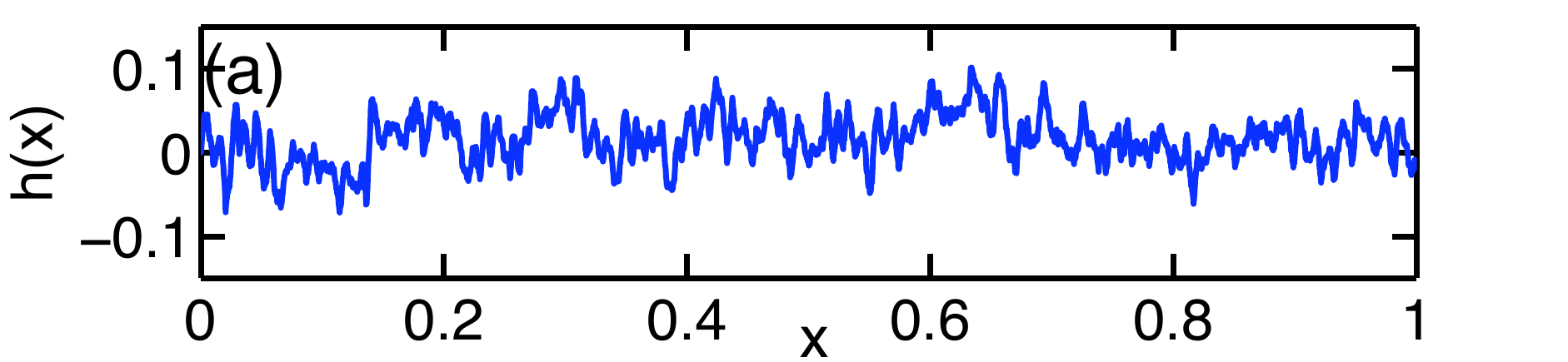}}
\centerline{\includegraphics[width=8cm]{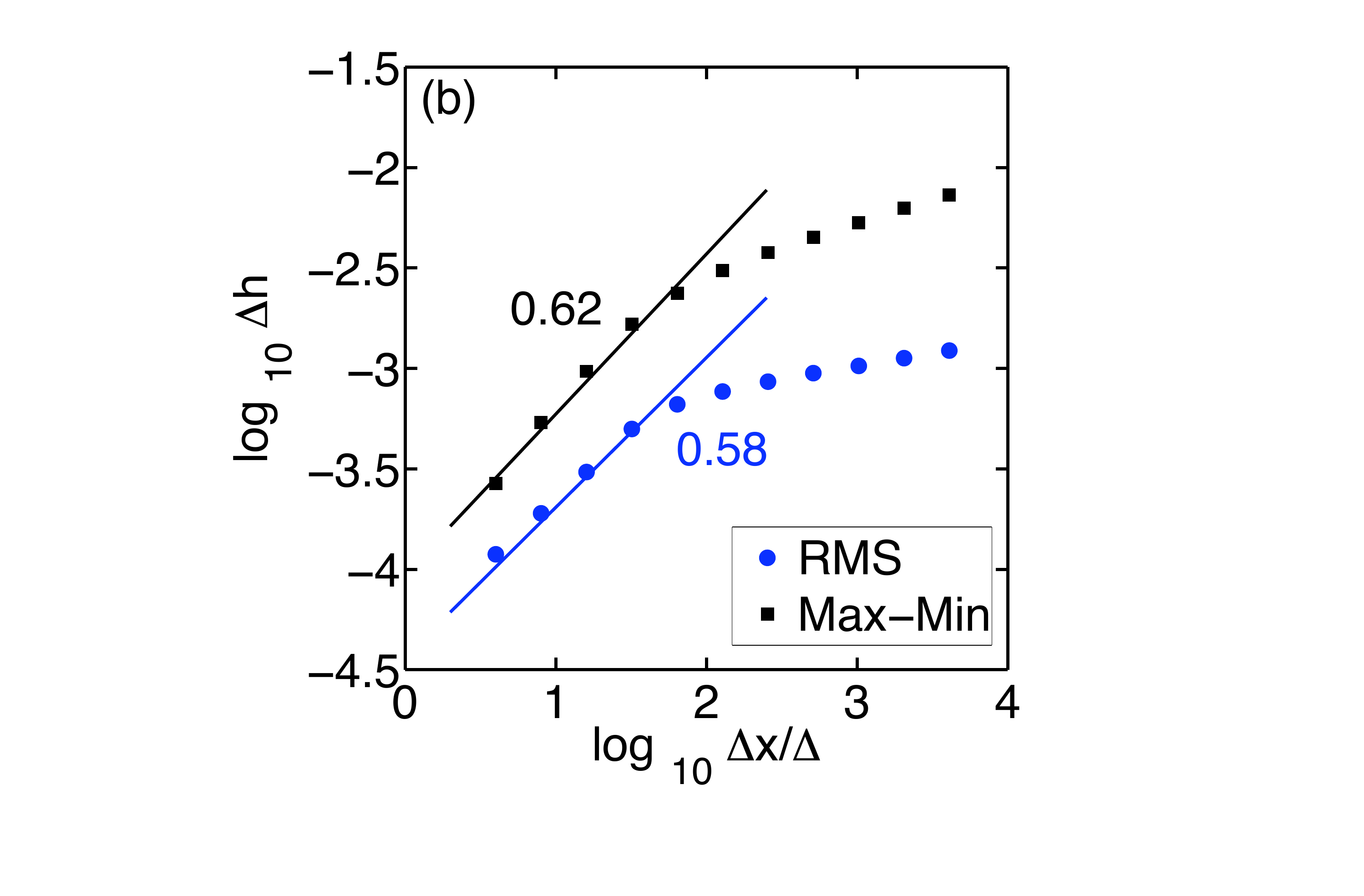} \includegraphics[width=8cm]{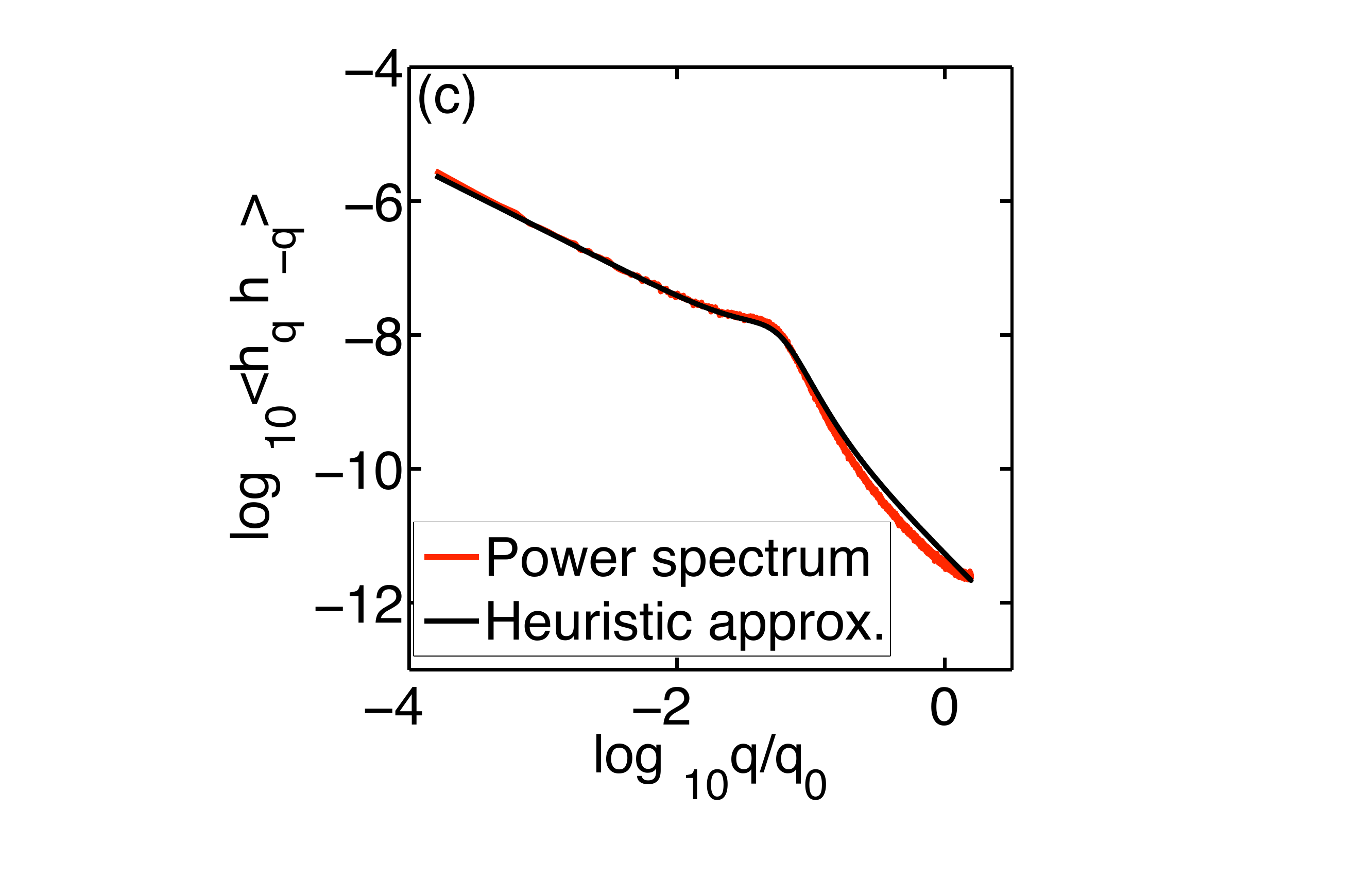}}
\caption{(Color online) (a) A crack path simulated using $\Delta=10^{-4}$, $D_1=4\times10^{-3}$ and $\beta=-64$. (b) RMS and Min-Max curves, the fit is just a guide to the eye (c) The power spectrum of the crack path over $100$ realizations and a theoretical prediction based on the heuristic approximation given by Eq.~(\ref{eq:h3-cPS}).
 \label{fig:roughT}}
\end{figure}

Let us now study of the effect of the $T$-stress term in Eq.~(\ref{eq:start}). Still, we consider only the stable regime, that is when $\beta<0$. In real experimental situations, one expects $\beta\simeq -1$ and most physical systems can be well described by the case $\beta=0$. Indeed our analysis and numerical results show that Eq.~(\ref{eq:start}) with $\beta\simeq -1$ exhibits scaling behaviour that is very close to the case $\beta=0$. Nevertheless, in order to demonstrate the impact of a non-zero $T$-stress, a large value of $\beta$ is used. Fig.~\ref{fig:roughT} shows results of crack paths grown with $\beta=-64$ and the corresponding power spectrum.

There are no analytical results for $\beta\neq0$. However, one expects a simplification in the spirit of the previous sections, i.e that averaging over $\eta_1$ would yield an effective Langevin equation of the form
\begin{equation}
 h''(x) = -\left[ C_1 h'(x) + C_2 h(x) - C_3 \beta \int_{-\infty}^x \frac{h'(x')}{\sqrt{x-x'}} dx' \right] + \eta_2'(x)
 \label{eq:h3-c} \ ,
\end{equation}
where $C_1,C_2$ and $C_3$ are renormalized deterministic prefactors, that would in general depend on $D_1, \Delta$ and $\beta$. The spectrum of the solution averaged over $\eta_2$ would then be given by
\begin{equation}
 \left \langle h_q h_{-q} \right \rangle_{\eta_1,\eta_2} = \frac{D_2 q^2 + \theta_0^2}{\left( q^2 - C_2 - C_3\sqrt{\frac{\pi}{2}}\beta \sqrt q \right)^2  + \left(C_3\sqrt{\frac{\pi}{2}}\beta \sqrt q  + C_1 q \right)^2}
 \label{eq:h3-cPS} \ .
\end{equation}
By comparing with the numerical results we find evidence indicating that one needs to take $C_1=C_2=C_3=C-\beta$, where $C\equiv D_1/\Delta$ is defined similarly to the case $\beta=0$. Using this modified form, the agreement with the simulation is good (see Fig.~\ref{fig:roughT}c). Note that in the expression (\ref{eq:h3-c}) the term related to $\beta$ is not dominant for small $q$'s  neither for large $q$'s which means that it does not modify the shape of the spectrum in a dramatic way. This is consistent with the power-counting argument mentioned before.

\begin{figure}[ht]
\centerline{\includegraphics[width=8cm]{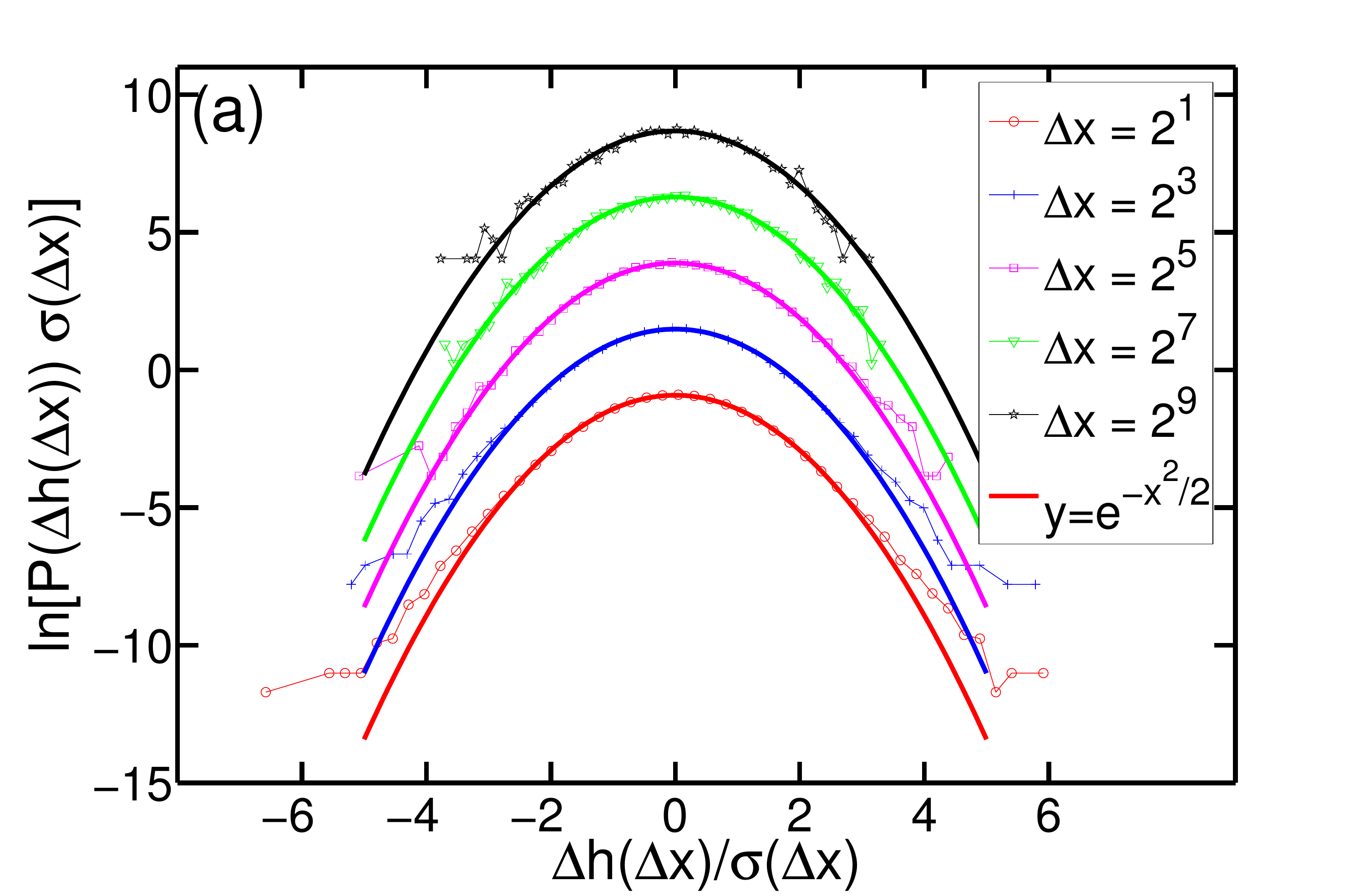}
\includegraphics[width=8cm]{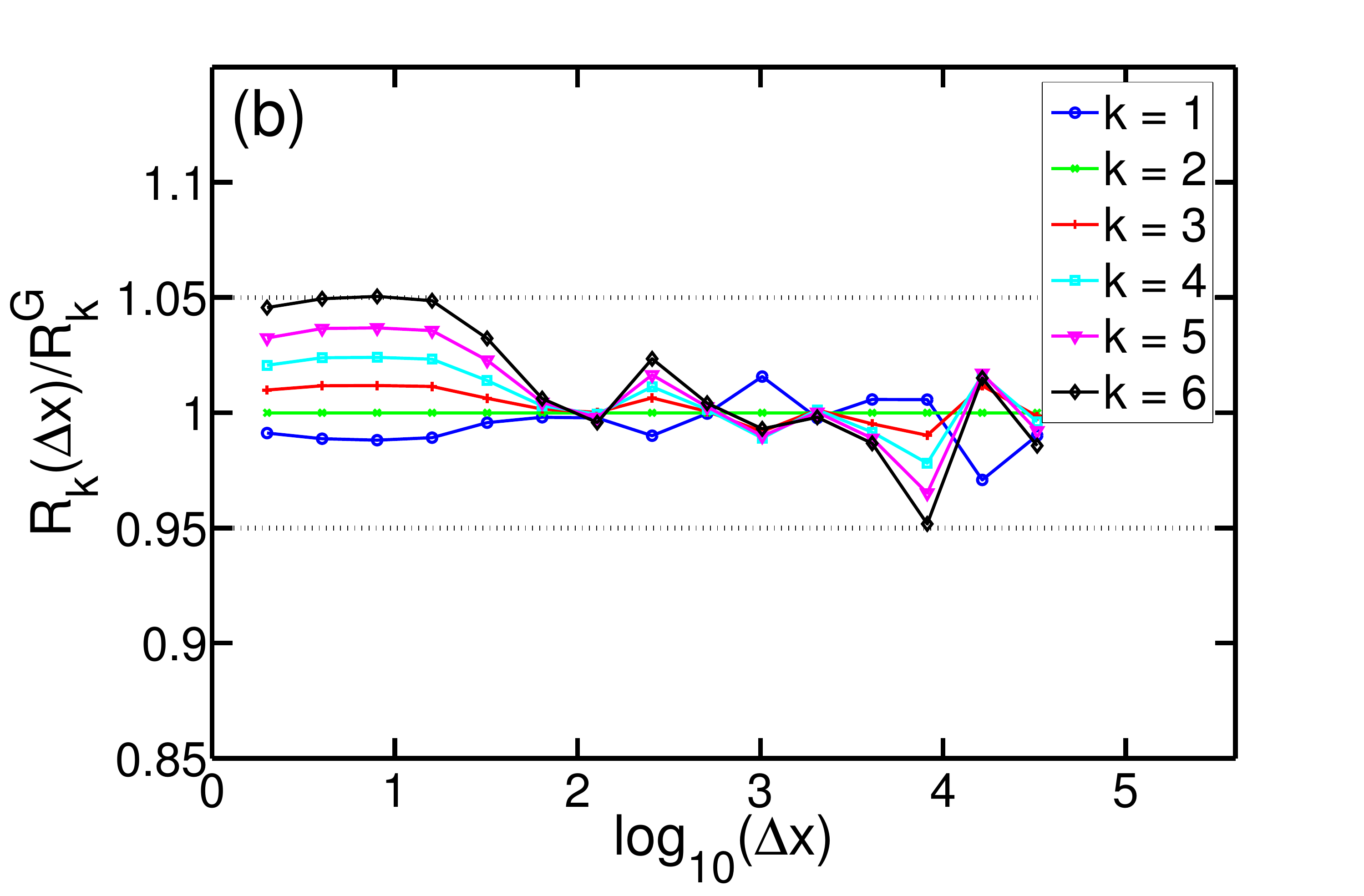}}
\centerline{\includegraphics[width=8cm]{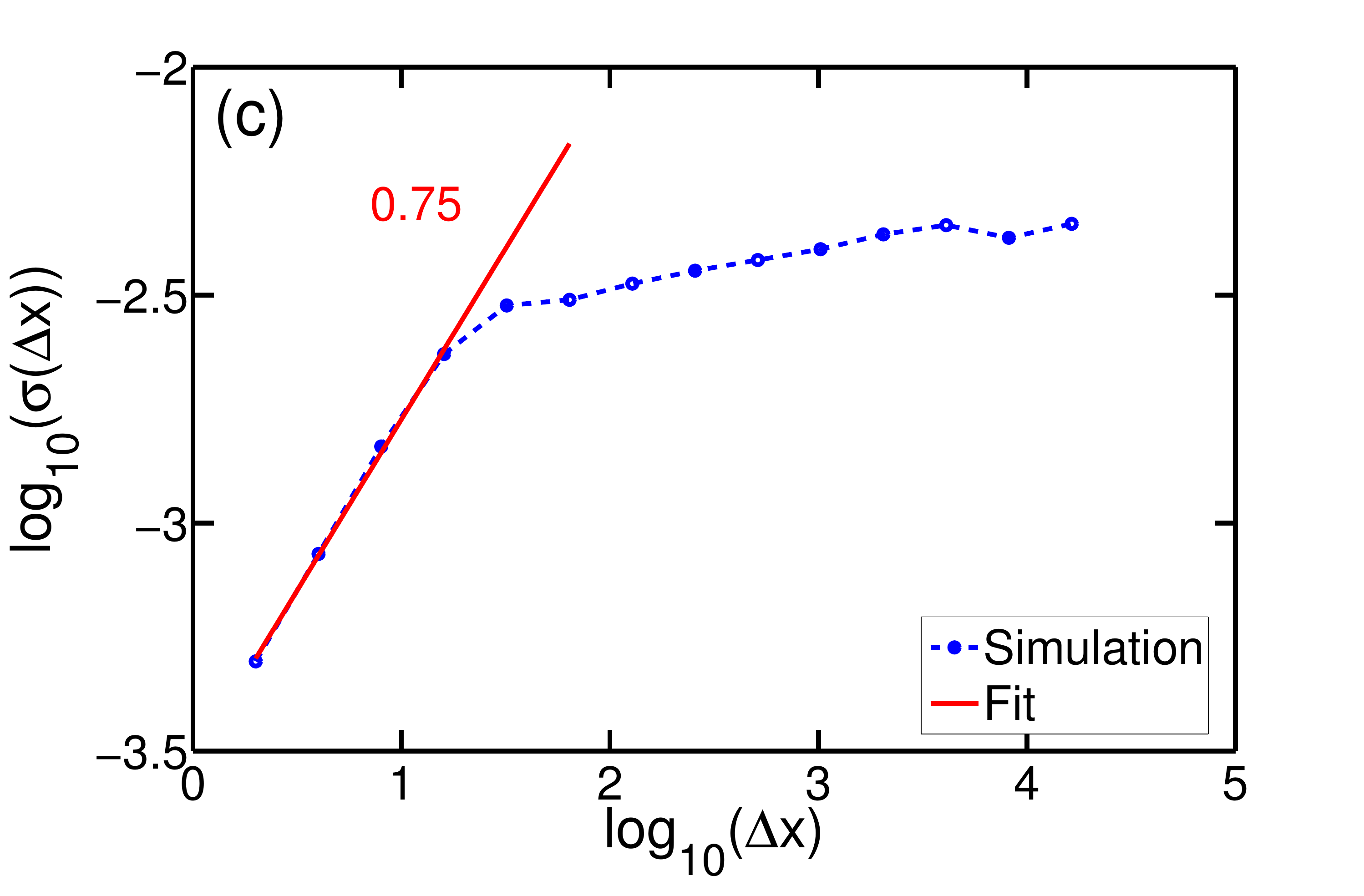}}
\caption{(Color online) (a) The statistical distribution for height increments for crack paths simulated using $\Delta=10^{-4}$, $D_1=4\times10^{-3}$ and $\beta=-64$ on a semi-log scale. For each $\Delta x$ we plot in addition a pure Gaussian as a guide to the eye. Note that we shifted the various distributions logarithmically for visual clarity. (b) A comparison of the moments $R_k(\Delta x)$ to those expected for a Gaussian distribution $R_k^G$ for $k=1,\ldots ,6$, revealing some deviations from Gaussianity. The dashed lines marks a $5\%$ deviation interval. (c) The RMS $\sigma(\Delta x)$ obtained from the simulation on a log-log scale. The first part of the curve that seems to follow a power law is fitted and an exponent close to $0.75$ is obtained.
\label{fig:PDFT}}
\end{figure}

The probability distribution functions for height increments can be easily computed numerically. Fig.~\ref{fig:PDFT} reports the same details as for the case $T=0$. Figs.~\ref{fig:PDFT}a-\ref{fig:PDFT}b show deviations from Gaussianity at large scales meaning that the PDF of the height differences are not self-similar. Since the first requirement for self-affinity is violated, it is clear that an analysis of the scaling of the variance is biased, and the results it yields should be taken prudently. Fig.~\ref{fig:PDFT}c reports such an analysis, i.e. $\sigma(\Delta x)$, and interestingly shows that the deviations from Gaussianity are sufficient to produce similar artifacts to those seen in the previous sections, namely a seemingly power law behaviour for small values of $\Delta x$. We verified that the power spectrum of the crack paths does not manifest the same bias, but rather reproduces a $q^{-2}$ behaviour at the tail as expected from Eq.~(\ref{eq:h3-cPS}) (implying $\zeta=0.5$). Anyway, the fact that the crack paths becomes flat at the large scales is independent of the method used.

\section{On self-affinity of crack surfaces}

So far we discussed the stability of crack paths in heterogeneous media, and the possible shapes they can take. It turned out that in the stable regime ($\beta \le 0$) one can get many possible types of patterns depending on the parameters of the model, including random patterns that seem to resemble self similar shapes. Actually, the analysis whose results are summarized in Figs.~\ref{fig:0.6}b-\ref{fig:0.8}b supports this observation and suggests that not only can one produce self-similar shapes but also a large family of such shapes that seem to span a wide range of different roughness exponents.

However, the analysis of the corresponding power spectra of these crack paths yields a different conclusion. A power spectrum of the form (\ref{eq:h2-cPS}) or (\ref{eq:h3-cPS}) means that strictly speaking the shape is not self-affine. To be more precise, it implies the existence of a self-affine structure on a small scale described by a roughness of $\zeta=1/2$ (deduced from the large $q$ behaviour of the spectrum, i.e. $1/q^2$) which is super-imposed on a decaying function or on damped oscillations at larger scales, and thus the
spectra crosses over to a flat behaviour for small $q$'s (possibly with a peak at some particular $q_0$ as in Fig.~\ref{fig:eta1}b). In this context, an attempt to fit a straight line to the power spectrum for intermediate values of $q$ might yield a seemingly reasonable fit for one/two decades but is certainly unjustified.

So how does one settle the difference between the results that come form the real space and the Fourier space approaches? The answer to this question is not restricted to crack surfaces and is related to a general discussion of reliability of self-affine measurements. A starting point is the work of Schmittbuhl {\it et a}l.~\cite{Schmittbuhl95a} that reviews various methods to extract the roughness exponent from measured or simulated profiles. In~\cite{Schmittbuhl95a}, the authors compare between various methods with respect to different artifacts that can appear in the data, such as misorientation or signal amplification. Interestingly, they came up with a useful sensitivity assessment of each method with respect to the biases. However, they did not discuss the case when a self-affine structure is imposed on an oscillating background, or more generally on a bias which is not translation invariant. For this case, we claim that the real-space methods (such as Min-Max and RMS) are highly vulnerable, while the power-spectrum is, naturally, very robust. An extreme example is given below in Fig.~\ref{fig:sin}, where a pure sinusoidal path, which is clearly not a self-affine profile, is analyzed using the Min-Max and the RMS methods. The resulting curves misleadingly reveal $3$ decades of self-affine behaviour (actually, an arbitrary number of decades can be devised easily), with a roughness exponent $\zeta\simeq 0.98$. In contrast, a Fourier analysis of this profile gives essentially a delta function localized at the wavelength of the oscillations. This artifact has not been discussed in ~\cite{Schmittbuhl95a} nor in other reviews of existing methods for measuring the roughness exponent such as~\cite{RafiFrac,Oystein}.

\begin{figure}[ht]
\centerline{\includegraphics[width=8cm]{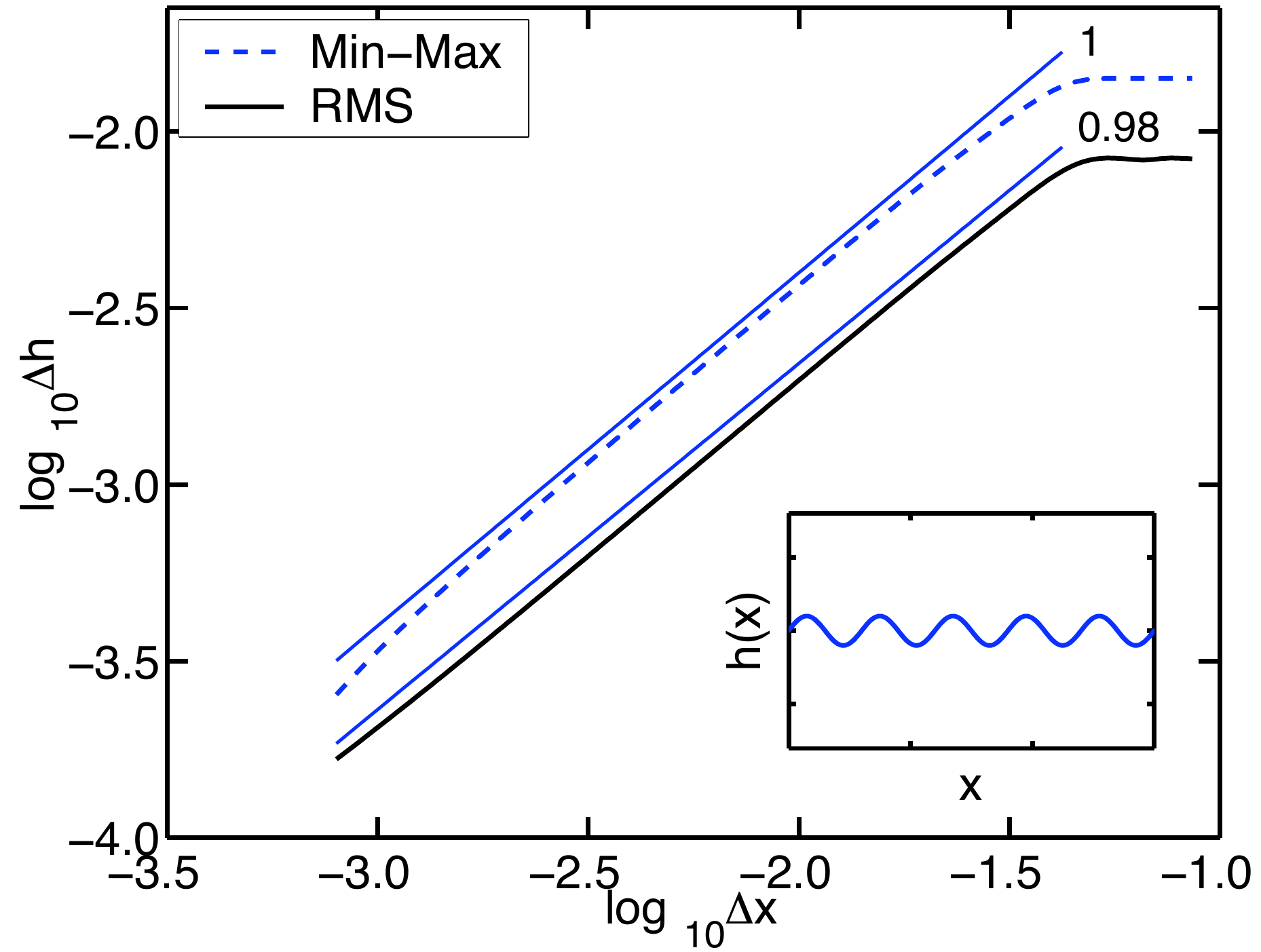}}
\caption{A pure sinusoidal path (inset) and the result of an RMS as well as a Min-Max
analysis of that path, "revealing" self-affinity over $3$ decades
with a roughness exponent $\zeta=0.98$.
\label{fig:sin}}
\end{figure}

Back to our model, this extreme example clearly favors Fourier based analysis over real-space approaches. Moreover, it can also explain why by varying the parameters of the model and by using a real-space analysis, we could obtain a whole range of $\zeta$'s between $0.5$ (which is the roughness exponent of a simple random walk) and $1$. This possibility is of course excluded when looking at the power spectrum where we have clear predictions for the its form.

Finally, let us discuss the approach of examining self-affinity through an analysis of the whole PDF of height differences $P(\Delta h(\Delta x))$. In principle, this method allows to check in a very precise way the self-affinity of crack paths and yields the roughness exponent. However, this method is not easy to implement since it needs a large amount of data that is typically much more than the amount of data that a usual experiment can yield. In our simulations we could use relatively long cracks of $10^7$ pixels with a reasonable effort. Even with this large amount of raw data it was not easy to get rid of artifacts in the PDF such as over estimates of the roughness exponent and detection of deviations from Gaussianity, while the power spectrum could perform better already with smaller samples. This means that while in principle the PDF method is superior to other ones, when discussing real samples which are always of a limited precision and resolution, other methods such as the analysis of the power spectrum (resulting only from a $2$-point statistics) are usually better.

\section{Conclusion and perspectives}

In this paper, we studied the stability and roughness of slow cracks in 2D disordered materials with respect to the disorder. Our approach relies on a solid ground in both mechanics and statistical physics. After proposing an equation of motion of a crack tip in a 2D disordered material, we first generalize the well known $T$-criterion predicted to disordered materials and then describe the roughening of crack paths. Using this equation of motion, we observe numerically various possible patterns, including oscillating, decaying and rough paths. We analyze the rough cracks using commonly used techniques. By using real-space methods we are able to obtain a whole range of roughness exponents, while the power-spectrum of the paths does not support these findings. Thanks to an exact result~\cite{EPL07} we are able to predict the power spectrum analytically for $T=0$ (Eq.~(\ref{eq:h2-cPS})), and approximately for $T<0$ (Eq.~(\ref{eq:h3-cPS})). These analytical results suggest that the shapes are not self-affine, but rather flat objects on the large scale and random-walk like objects (with $\zeta=0.5$) at small scales. We conclude that in such situations real-space methods are very vulnerable (as they mix the scales and can yield seemingly self-affinity with almost any $\zeta$ between $0.5$ and $1$), while the Fourier-space approach is much more appropriate and thus preferred. The last point is not only relevant to the analysis of cracks, but applies to analysis of self-affine shapes where an oscillating background is likely to exist. We hope that this will contribute to the general discussion of reliability of self-affine measurements.

From an experimental point of view, it could be interesting to use expressions like Eqs.~(\ref{eq:h2-cPS}) and (\ref{eq:h3-cPS}) to fit the experimental data of 2D rough cracks. An important prediction implied by these expressions is that the scale at which the crossover to a flat shape occurs is roughly $q \sim \sqrt{C}$, that is a geometric mean of a geometric length-scale and a disorder length-scale (which is the density of disorder $\rho$ times the amplitude of the toughness fluctuations $D_1$). This suggests that by varying the width of the strip in which the crack propagates and/or tuning the density and amplitude of the heterogeneities one can change the crossover scale. Another interesting prediction is that by varying the sign of the combination $C-4$ from positive to negative (for example by reducing the density of heterogeneities) one can switch between a pure exponential decay of the shear perturbations to an oscillatory one. In the presence of many shear perturbations, this property can be easily observed as a peak appearing in the power spectrum, as the transition to an oscillatory response occurs. It could be the case that some generalization of this effect is responsible to the transition reported in~\cite{Deegan03} from oscillatory to rough cracks. It could be that an extension of the approach developed here, together with the recent results of \cite{thermique} to account for thermal effects, provide a proper explanation for this transition.

An open question, not developed in this work, is how does the material micro-structure affect the roughness of the cracks paths. Here, we considered only short-ranged disorder, modeled by $\delta$-correlated noise for both kinds of disorder $D_1$ and $D_2$ (corresponding to toughness and shear fluctuations respectively). It is clear that considering off-lattice heterogeneities and taking into account the different correlation properties of $\eta_1$ and $\eta_2$ would make the paths and their resulting power spectra more complex - a well known phenomenon in stochastic systems \cite{Spatial,Temporal}. Furthermore, long-range power law correlations that are known to exist in certain materials such as quasi-crystals \cite{Lifshitz03}, porous materials \cite{Katzav06b} and others~\cite{materials} could yield long range correlations in the toughness/shear fluctuations, and thus lead to yet richer phenomena. Actually, a recent discrete numerical model of fracture has shown that long-range correlations in the disorder and its anisotropy can lead to non-universal scaling exponents~\cite{Mehdi}. These results may change dramatically future approaches to problems of crack propagation in disordered materials.

Finally, a fundamental open question is whether this work could help to understand the enigma of crack roughness in 3D samples. It is hoped that the in-plane roughness, mentioned above, and studied experimentally~\cite{Daguier95,Schmittbuhl97,Delaplace99,Maloy06,Chopin}, may provide an important starting point. More precisely, modeling the simplified 3D problem \cite{Gao89,Ramanathan97,Ramanathan98,Adda06,Katzav06a,wetting} may be combined with the out-of-the-plane fluctuations (namely the current work) to yield a full 3D theory. Various elements needed in that direction can be found in the literature~\cite{Pindra10,Vasoya13,Ramanathan97b,Ball95,Ball97} but such a theory is still far from being formulated.

\section*{Acknowledgments}

We thank B. Derrida, A. Boudaoud, K. J. M\aa l\o y and S. Santucci for fruitful discussions.


\begin{appendix}
\section{Measurements of self-affine surfaces}

For any random surface, one of the most studied quantities to characterize its geometry is the roughness exponent. For one dimensional crack paths embedded in two dimensional materials one single exponent $\zeta$ is needed. This exponent is also known as the Hurst exponent. There are many methods to measure self-affine exponents~\cite{Schmittbuhl95a,RafiFrac,Oystein} and we will shortly review here four that are commonly used.

Let us parameterize the path using the function $h(x)$ with $0 \le x \le L$. Using the variable bandwidth methods~\cite{Schmittbuhl95a,RafiFrac,Oystein}, the roughness exponent can be related to the scaling of the width in the $h$-direction as a function of a window size in the $x$-direction. More specifically one expects $\Delta h \sim (\Delta x)^\zeta$. What
is left is to specify a way to define the width $\Delta h$. A possible definition is given by the standard deviation (or RMS) of the height profile
\begin{equation}
\Delta h_{\text{RMS}} \left(\Delta x\right) \equiv \sqrt{\frac{1}{L}
\int_0^L \left[( h(x+\Delta x) - h(x) \right]^2 dx} \sim
\left(\Delta x\right)^{\zeta}
 \label{scaling1}\ .
\end{equation}
We will refer to it as the RMS (Root Mean Square) method. Another variable bandwidth method is the Min-Max method defined by
\begin{equation}
\Delta h_{\text{Min-Max}} \left(\Delta x\right) \equiv \frac{1}{L} \int_0^L
\left| \max \{h(x')\}_{x<x'<x+\Delta x} - \min
\{h(x')\}_{x<x'<x+\Delta x} \right| dx \sim \left(\Delta
x\right)^{\zeta}
 \label{scaling2}.
\end{equation}
A more elaborated way to extract the roughness exponent, which allows to test the self-affinity at the same time~\cite{Santucci07,Auradou07,Horst09,Bonamy09}, is based on examining directly the Probability Distribution Function (PDF) of the discrete gradient
\begin{equation}
\Delta h (\Delta x) \equiv h(x+\Delta x) - h(x) \label{eq:delh}
\end{equation}
rather than just looking at its second moment as in the RMS method. In order to implement this method, one needs to plot the (properly normalized) PDF of $\Delta h$ for every $\Delta x$. If the shape is self-affine two conditions should be met. First, the PDFs emanating from different values of $\Delta x$, namely $P\left(\Delta h (\Delta x)\right)$, should collapse on to a single curve. Second, the normalization factor of the various PDFs should scale as $(\Delta x)^\zeta$. These requirements can be summarized by
\begin{equation}
P\left(\Delta h (\Delta x)\right) = \lambda^\zeta
P\left(\lambda^{-\zeta} \Delta h (\lambda \Delta x)\right) \, ,
 \label{scaling4}
\end{equation}
where $\lambda$ is the scale factor. An important remark is that the master curve for the PDF $P(\Delta h)$ does not have to be Gaussian (although it can be) in order to imply self-affinity, it just has to be the same across all the different scales $\Delta x$. Although this is a very rigorous and precise method that allows both testing for self-affinity and determining the roughness exponent, it is not always possible to implement it in experimental systems as it requires precise data that spans many orders of magnitude.

The last method that we will describe is not based on real-space measurements, but rather on the Fourier transform of the path, denoted here as $h_q$, where $q$ is the wave-number. More precisely, the power-spectrum $C(q)$ of a self-affine path is expected to scale as
\begin{equation}
C(q) \equiv \left\langle h_q h_{-q} \right\rangle \sim q^{-1-2\zeta}
 \label{scaling3}.
\end{equation}
\end{appendix}

\end{document}